\documentclass[preprint,showpacs,preprintnumbers,amsmath,amssymb]{revtex4}

\usepackage{graphicx}
\usepackage{dcolumn}
\usepackage{bm}
\usepackage{amsmath,dsfont}
\usepackage{amssymb,amsthm,amscd, amsbsy, array}
\usepackage{graphics,graphicx,xcolor}
\usepackage{etex}
\usepackage[all]{xy}

\numberwithin{equation}{section}

\newcommand{\half}{{\scriptstyle{\frac{1}{2}}}}
\def\2{{\half}}
\newcommand{\const}{\mathop{\rm const}\nolimits}

\def\va{{\bm{a}}}
\def\vc{{\bm{c}}}

\newcommand{\vp}{{\bm p}}
\def\vE{{\bm E}}

\def\beq{\begin{equation}}
\def\eeq{\end{equation}}
\def\beqa{\begin{eqnarray}}
\def\eeqa{\end{eqnarray}}
\def\nn{\nonumber}
\def\barray{\left(\begin{array}}
\def\earray{\end{array}\right)}
\def\barraynb{\begin{array}}
\def\earraynb{\end{array}}

\def\smallover#1/#2{\hbox{$\textstyle\frac{#1}{#2}$}} %

\def\vx{{\bm{x}}}

\def\vX{{\bm{X}}}
\def\vY{{\bm{Y}}}

\def\cK{{\cal K}}
\def\cJ{{\cal J}}
\def\cP{{\cal P}}
\newcommand{\vcK}{{\bm \cK}}

\begin{document}


\title{
Chiral decomposition in the non-commutative Landau problem
\\[6pt]
}

\author{P-M. Zhang}\email{zhpm-at-impcas.ac.cn}

\author{P.~A.~Horvathy\footnote{Permanent address~:
{\it Laboratoire de Math\'ematiques et de Physique
Th\'eorique}, Tours University
(France).}}\email{ horvathy-at-lmpt.univ-tours.fr}
\affiliation{Institute of Modern Physics, Chinese Academy of Sciences
\\
Lanzhou (China)
}
\date{\today}

\begin{abstract}
The decomposition  of the non-commutative Landau (NCL) system  into two uncoupled one-dimensional chiral components, advocated by Alvarez, Gomis,
Kamimura and Plyushchay \cite{AGKP}, is
generalized to nonvanishing electric fields.
This allows us to discuss the main properties of the NCL problem including its exotic Newton-Hooke symmetry and its relation to the Hall effect.
The ``phase transition''  when
the magnetic field crosses a critical value determined by the non-commutative parameter is studied in detail.
\end{abstract}

\pacs{04.50.Cd,11.30.-j,02.20.Sv\\
Annals of Physics [in press] (2012)
DOI information: 10.1016/j.aop.2012.02.014}

\maketitle

\newpage
\null\newpage
\tableofcontents
\newpage

\section{Introduction}

In two remarkable papers  Alvarez, Gomis, Kamimura and Plyushchay  \cite{AGKP}
pointed out that two uncoupled  $1$d chiral oscillators yield, when combined, an interesting non-commutative system in the plane.
The purely-magnetic ``exotic'' Landau problem (NCLP) \cite{DHexo,NCLandau} can, in particular, be obtained for a suitable choice of the parameters.
The inclusion of the electric field which, for non-commutative particles, also induces an anomalous velocity term
(missed in Ref. \cite{AGKP}), is important, though, in the study of a Bloch electron
 \cite{Bloch}, and for both
the ordinary
and  the anomalous Hall effects \cite{QHE,AHE},
for example.

In this paper we generalize
the chiral oscillator --- exotic Landau-problem correspondence
 to non-vanishing electric fields.
 The examples of a constant and of a harmonic
electric field are  worked out explicitly. We also re-derive, along the lines indicated by Ref. \cite{AGKP}, the recently found ``exotic'' Newton -- Hooke symmetry \cite{ZH-II,NewtonHooke}.

The chiral framework is particularly useful for quantization which can be achieved, just like classically \cite{ZH-II}, using conserved quantities
alone.

The system becomes singular when the magnetic field
takes a certain critical value $B_c$\,; the
physical meaning of the ``phase transition'' when this critical value
 is crossed is studied in detail.

It is worth mentioning that chiral oscillators were used before in
 the ordinary Landau problem and used to
 explain the non-commutativity of the guiding center coordinates  \cite{Banerjee, Sivasubramanian}.

\goodbreak
\section{Exotic dynamics}\label{exotic}

Let us first summarize some of the main features of ``exotic'' particles.
Such a particle,  moving in a planar electromagnetic field  $\vE,B$,
[assumed static for simplicity]
is described by the equations
\cite{DHexo,NCLandau},
\beqa
m^*
\dot{x}^i=p^i-me\theta\varepsilon^{ij}E^j,
\qquad
\dot{p}^i=eB\varepsilon^{ij}\dot{x}^j+eE^i,
\label{exoeqmot}
\eeqa
where  $m,\,e$ and $\theta$ are the
mass, charge and non-commutative parameter,
respectively, and
$m^*=m(1-eB\theta)$ is the effective mass.
Note also,
in the first relation, the ``anomalous velocity term'',
which is only absent for either in the purely magnetic-
($\vE=0$) or in  the commutative ($\theta=0$) case, and is indeed responsible for
the Hall behavior \cite{DHexo,NCLandau,AHE}.

It has been proved recently that when the charges, masses, and non-commutative parameters
of a collection of exotic particles satisfy the  generalized Kohn conditions
$e_a/m_a=\const.$ and $e_a\theta_a=\const.$, then
the system of interacting ``exotic'' electrons splits into
internal and center-of-mass motions, with the
latter behaving as a single exotic particle \cite{ZH-II}.
It is enough to study one particle therefore.

Eqns. (\ref{exoeqmot}) derive from the symplectic form and Hamiltonian,
\beqa
\Omega=
dp^i\wedge dx^i+\frac{\theta}{2}\varepsilon^{ij}dp^i\wedge dp^j+\frac{eB}{2}\varepsilon^{ij}dx^i\wedge dx^j,
\qquad
H=\frac{\vp^2}{2m}+V(\vx),
\eeqa
respectively, through the ``exotic'' Poisson brackets
\beqa
\{x^i,x^j\}=\frac{\theta}{1-eB\theta}\,\varepsilon^{ij},
\qquad
\{x^i,p^j\}=\frac{\delta^{ij}}{1-eB\theta},
\qquad
\{p^i,p^j\}=\frac{eB}{1-eB\theta}\,\varepsilon^{ij}.
\label{exoPB}
\eeqa
 When the magnetic field takes  the critical value
\beq
B=B_c=\frac{1}{e\theta}\,,
\label{critB}
\eeq
the system becomes singular~: the determinant
of the symplectic matrix is det $(\Omega_{\alpha\beta})=(m^*/m)^2=0$, and
consistency requires the Hall law,
\beq
p^i=me\theta\varepsilon^{ij}E^j,
\qquad
\dot{x}^i=\varepsilon^{ij}\frac{E^j}{B}\,,
\label{Hallaw}
\eeq
to be satisfied \cite{DHexo,NCLandau}.
This point will be further developed in  Sect.
\ref{critical}.

\goodbreak

\section{Splitting into chiral components}\label{Chiral}

Turning to our promised chiral decomposition,
 we only consider two particular cases,
since we are not yet able to tackle the problem for an arbitrary potential.

\subsection{Constant force}

Following Ref. \cite{AGKP}, let us now introduce chiral coordinates ${\vX}_{\pm}$ on
$4$d phase space,
\beq
p^i=eB\varepsilon^{ij}X^j_-,
\qquad
x^i=X_+^i+X_-^i\,.
\label{CEchircoord}
\eeq
In these terms,
\beqa
\Omega&=&\Omega_++\Omega_-=
\Big\{\frac{eB}{2}\left(\varepsilon^{ij}dX_+^i\wedge dX_+^j\right)\Big\}\;-
\;\,\Big\{
(1-eB\theta)\frac{eB}{2}\left(\varepsilon^{ij}dX_-^i\wedge dX_-^j\right)\Big\},\qquad
\label{EsplitOmega}
\\[8pt]
H&=&H_++H_-=\Big\{-e\vE\cdot\vX_+\Big\}\hskip17mm +\;\;
\Big\{\frac{(eB)^2}{2m}\vX_-^2-e\vE\cdot\vX_-\Big\}.
\label{EsplitHam}
\eeqa
The system splits therefore into two uncoupled systems, both with
$2$d phase spaces, described by respective coordinates
$\vX_{\pm}$. Note the negative sign in $\Omega_-$, justifying the expression ``chiral''.

Off the critical case, $eB\theta\neq1$, (\ref{EsplitOmega}) implies the fundamental Poisson Brackets
\beqa
\{X_+^i,X_+^j\}=-\frac{1}{eB}\,\varepsilon^{ij},
\qquad
\{X_-^i,X_+^j\}=0,
\qquad
\{X_-^i,X_-^j\}=\frac{1}{eB(1-eB\theta)}\,\varepsilon^{ij}.
\label{exoEPB}
\eeqa
The chiral equations of motion,
\beqa
\dot{X}_+^i=\varepsilon^{ij}\frac{E^j}{B}\,,
\qquad
m^*\dot{X}_-^i=eB\varepsilon^{ij}\,X_-^j
-m\,\varepsilon^{ij}\frac{E^j}{B}\,,
\label{chireqmot}
\eeqa
are solved as
\beqa
X_+^i(t)&=&\varepsilon^{ij}\frac{E^j}{B}\,t+X_+^i(0),
\label{X+mot}
\\[6pt]
X_-^i(t)&=&\Big[R\left(-\omega^{*}t\right)\big(\vX_{-}(0)-\vY\big)\Big]^i+Y^i,
\;\;
\vY=\frac{m\vE}{eB^2}=\const,
\;\;
\omega^*=\frac{eB}{m^*}.\quad
\label{X-mot}
\eeqa
 $X_+(t)$, physically the guiding center,
drifts therefore perpendicularly to the electric field  according to the Hall law,
while $\vX_-$ rotates with angular velocity
$(-\omega^*)$. Then a look at (\ref{CEchircoord}) allows us to conclude that, off
the critical case, our exotic particle
  moves  like an ordinary charged particle but with modified cyclotron frequency, $\omega^*=eB/m^*$,
where $m^*=m(1-eB\theta)$ is the critical mass.
Let us observe that
the body of the trajectory does not depend on $\theta$,
only its speed.

In the critical case
$
B=B_c=(e\theta)^{-1}
$
cf. (\ref{critB}),
 the second term in the
symplectic form (\ref{EsplitOmega}) is turned off, and the
$\vX_-$-degrees of freedom are lost. The second eqn. in
(\ref{chireqmot}) can nevertheless be satisfied
provided the initial condition is
$\vX_-(0)=\vY$. Then
 $\vX_-$ remains frozen into the fixed value
$
\vX_-(t)=\vY,
$
leaving us with
the $\vX_+$-equation alone in (\ref{chireqmot}), solved as
in (\ref{X+mot}). Then from (\ref{CEchircoord}) we infer that for  vanishing $m^*$ the ``whirling'' is eliminated and the motion is along straight lines according to the Hall law,
\beq
x^i(t)=X_+^i(t)+Y^i=
\varepsilon^{ij}\frac{E^j}{B}\,t+x^i(0),
\label{Hall}
\eeq
materializing the guiding center motion. The latter follows parallel lines
with the same velocity which can start from any point of the plane,
see FIG. \ref{BE3Dsweep}.

The  behavior when the magnetic field sweeps from very weak
to very strong crossing  the critical value $B_c$ is further
studied in Section \ref{critical}.

\subsection{Harmonic trap}

The splitting (\ref{CEchircoord}) works for the purely magnetic case,
but fails when we consider a harmonic trap with
spring constant $k\neq0$.
Generalizing (\ref{CEchircoord}) as
\beq
p^i=\epsilon^{ij}\big(\omega_+X_+^j+\omega_-X_-^j),
\qquad
\vx=\vX_++\vX_-\,,
\label{Ochircoord}
\eeq
with $\omega_\pm$ to be determined, we find
\begin{eqnarray}
\Omega&=&\underbrace{
(eB-2\omega_++\omega_+^2\theta)}_{\mu_+}
\,\frac{1}{2}
\varepsilon^{ij}dX_+^i\wedge dX_+^j
\;+\;\underbrace{
(eB-2\omega_-+\omega_-^2\theta)}_{\mu_-}
\,\frac{1}{2}\varepsilon^{ij}dX_-^i\wedge dX_-^j
\nn
\\[8pt]
&&+
\Big(eB+\omega_+\omega_-\theta-(\omega_{+}+\omega_{-})\Big)\,
\varepsilon^{ij}\big(dX_+^i\wedge dX_-^j\big),
\label{OSdec}
\end{eqnarray}
which splits into $\Omega_++\Omega_-$ when
the coefficient of the cross term vanishes,
\beq
eB+\omega_+\omega_-\theta-(\omega_{+}+\omega_{-})=0.
\eeq
For the Hamiltonian we find instead
\begin{eqnarray*}
H=\frac{\vp^2}{2m}+k\frac{\vx^2}{2}=
\underbrace{
\frac{1}{2m}\big(mk+\omega_+^2\big)\vX_+^2}_{H_{+}}
\;+\;\underbrace{
\frac{1}{2m}\big(mk+\omega_-^2\big)\vX_-^2}_{H_{-}}
+
\Big(k+\frac{\omega_+\omega_-}{m}\Big)\vX_+\cdot\vX_-\,,
\end{eqnarray*}
which splits as $H=H_++H_-$ when
\beq
\omega_+\omega_-+mk=0.
\eeq
Solving for $\omega_\pm$ allows us to deduce that choosing
\beq
\omega_\mp=\frac{eB-mk\theta\pm\sqrt{(eB-mk\theta)^2+4mk}}{2}
\eeq
separates the exotic Landau problem with harmonic potential
into two chiral oscillators.

In the purely magnetic case $k=0$ the previous choice
(\ref{EsplitOmega}) is recovered. Our formulae are also consistent with those  in the ordinary (commutative) Landau problem \cite{Banerjee,Sivasubramanian}.

Off the critical case, $eB\theta\neq1$,  the  Poisson Brackets read
\beq
\{X_+^i,X_+^j\}=-\frac{1}{\,\mu_+}\,\varepsilon^{ij},
\qquad
\{X_-^i,X_+^j\}=0,
\qquad
\{X_-^i,X_-^j\}=-\frac{1}{\,\mu_-}\,\varepsilon^{ij}
\eeq
with $\mu_\pm$ those coefficients
in (\ref{OSdec}),
yielding separated equations of motion,
\beq
m\mu_\pm\dot{X}_{\pm}^i=-(mk+\omega_\pm^2)\varepsilon^{ij}{X}_{\pm}^j.
\label{pmOsceqmot}
\eeq

\begin{figure}
\begin{center}
\includegraphics[scale=.17]{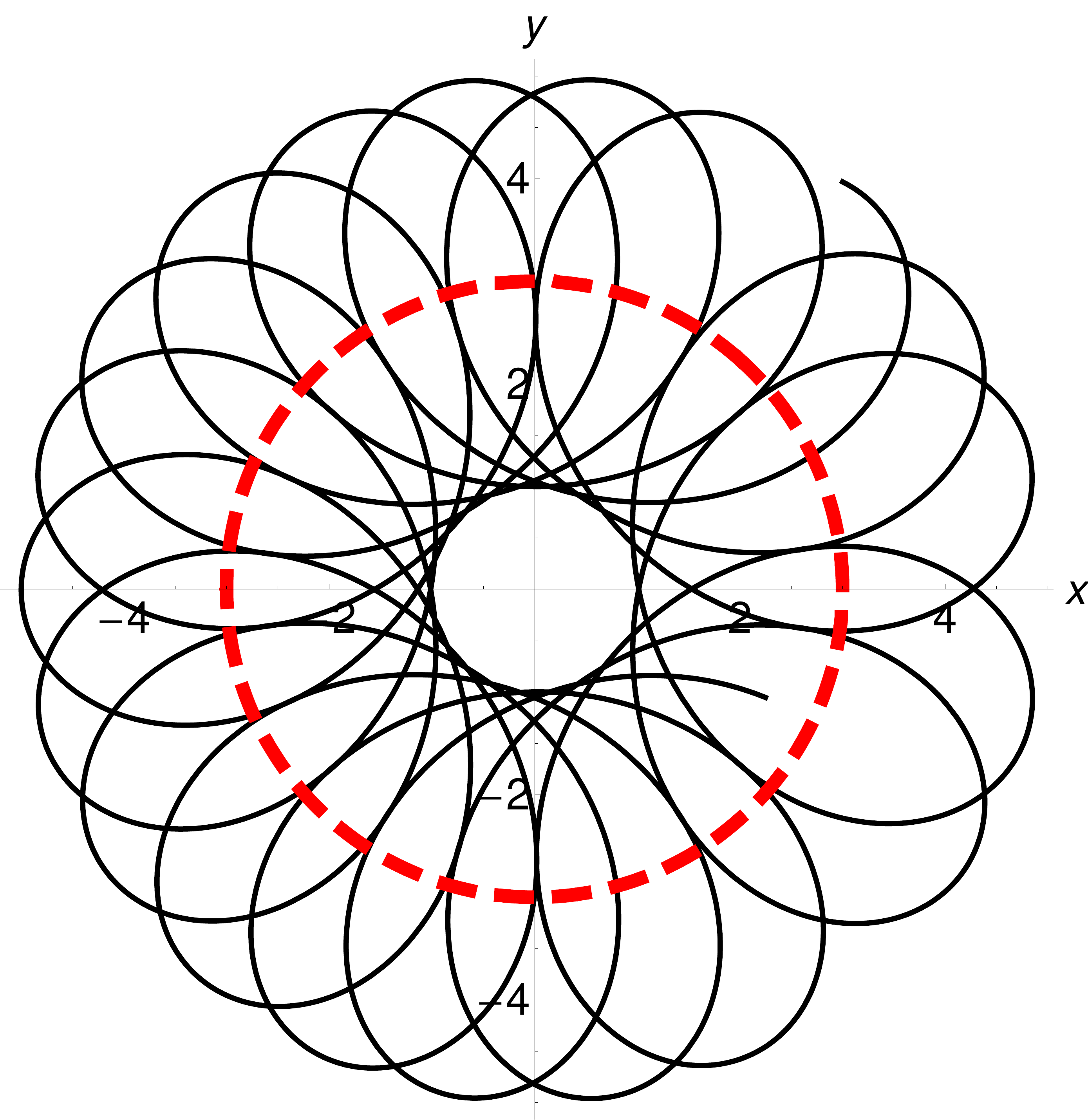}\quad
\includegraphics[scale=.17]{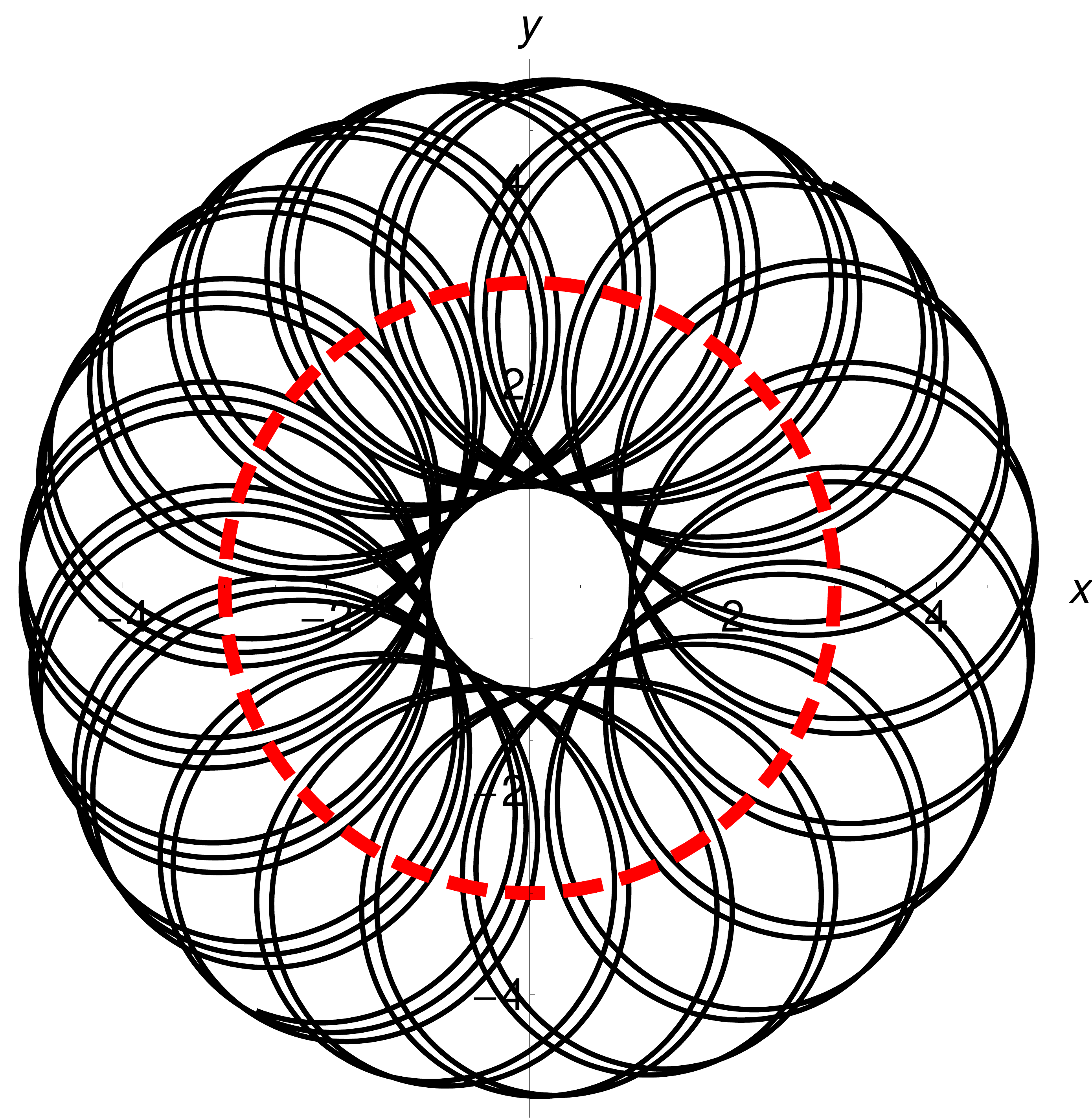}\\[8pt]
\includegraphics[scale=.17]{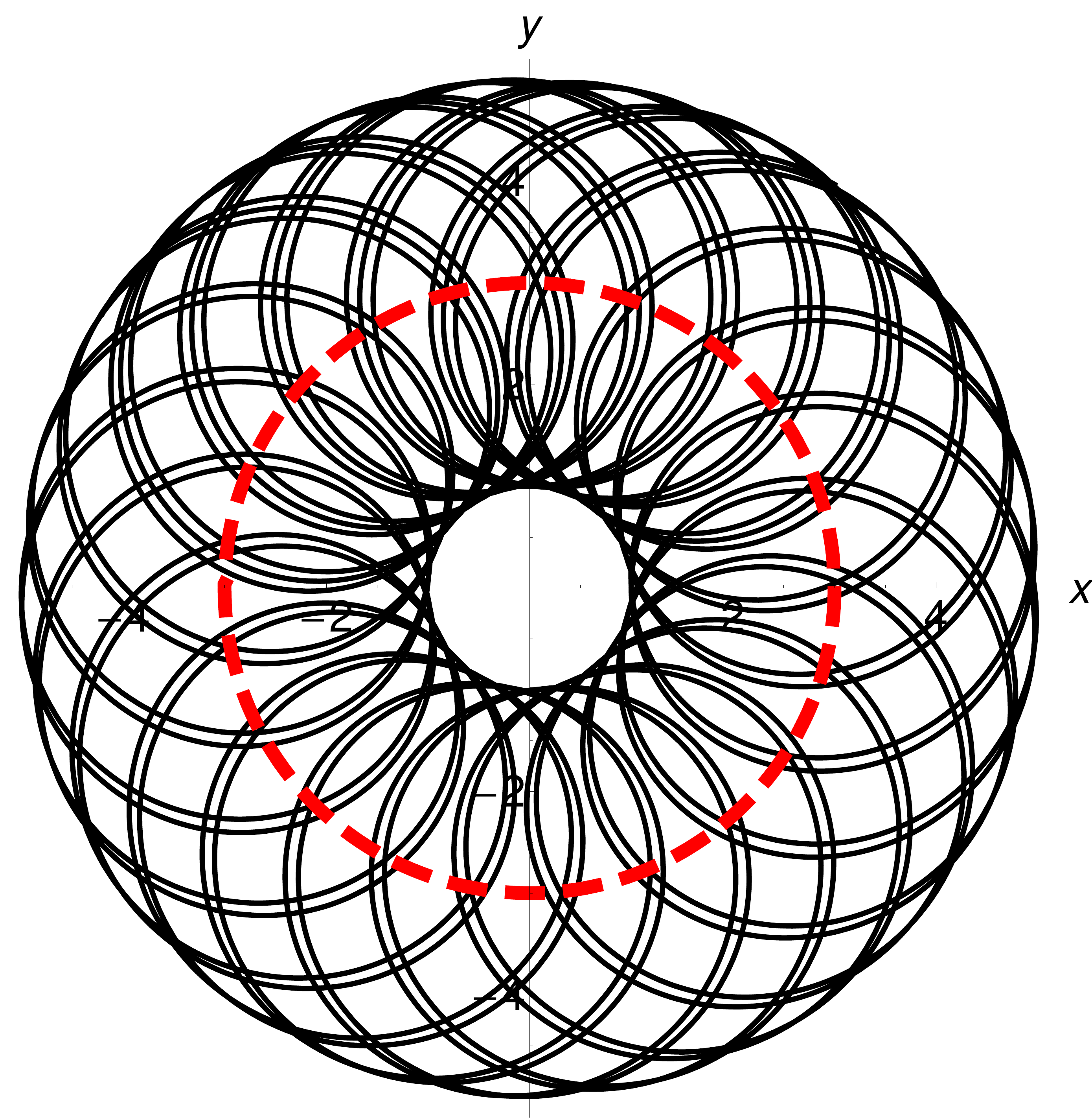}\quad
\includegraphics[scale=.17]{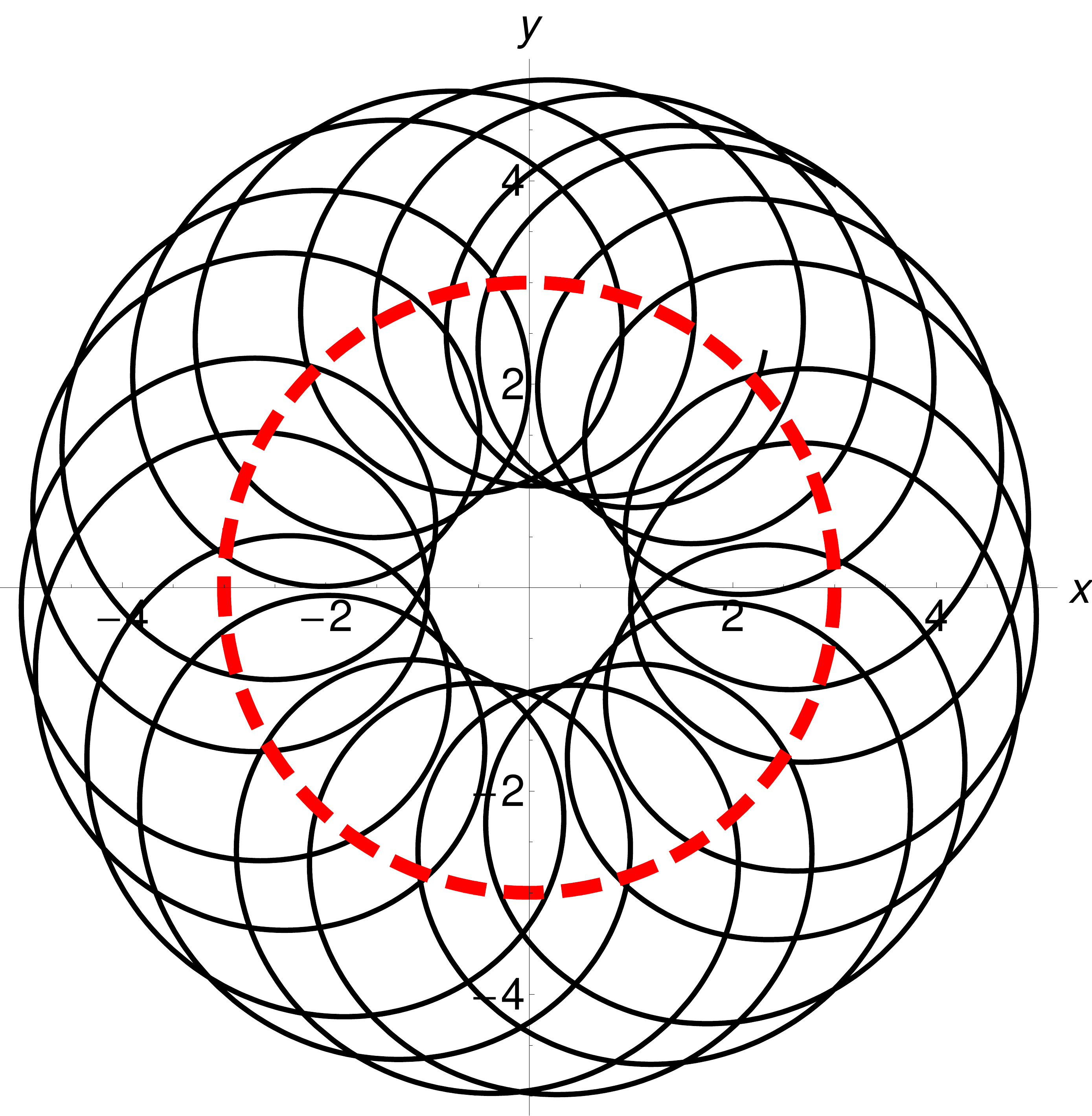}
\vspace{-8mm}
\end{center}
\caption{\it Trajectories in the
exotic Landau problem with an oscillator with $eB\theta$ sweeping through the range of parameters
 $eB\theta= 0.5,\,0.8,\,1.1,\,1.5$
and with initial conditions
$X^1_+(0)=2;\, X^2_+(0)=\sqrt 5; \, X^1_-(0)=1;\,
X^2_-(0)=\sqrt 3$, respectively. The dotted circle in the middle indicates the trajectory of the guiding center.
}
\label{BOFigure1-b}
\end{figure}

\begin{figure}
\begin{center}
\includegraphics[scale=.15]{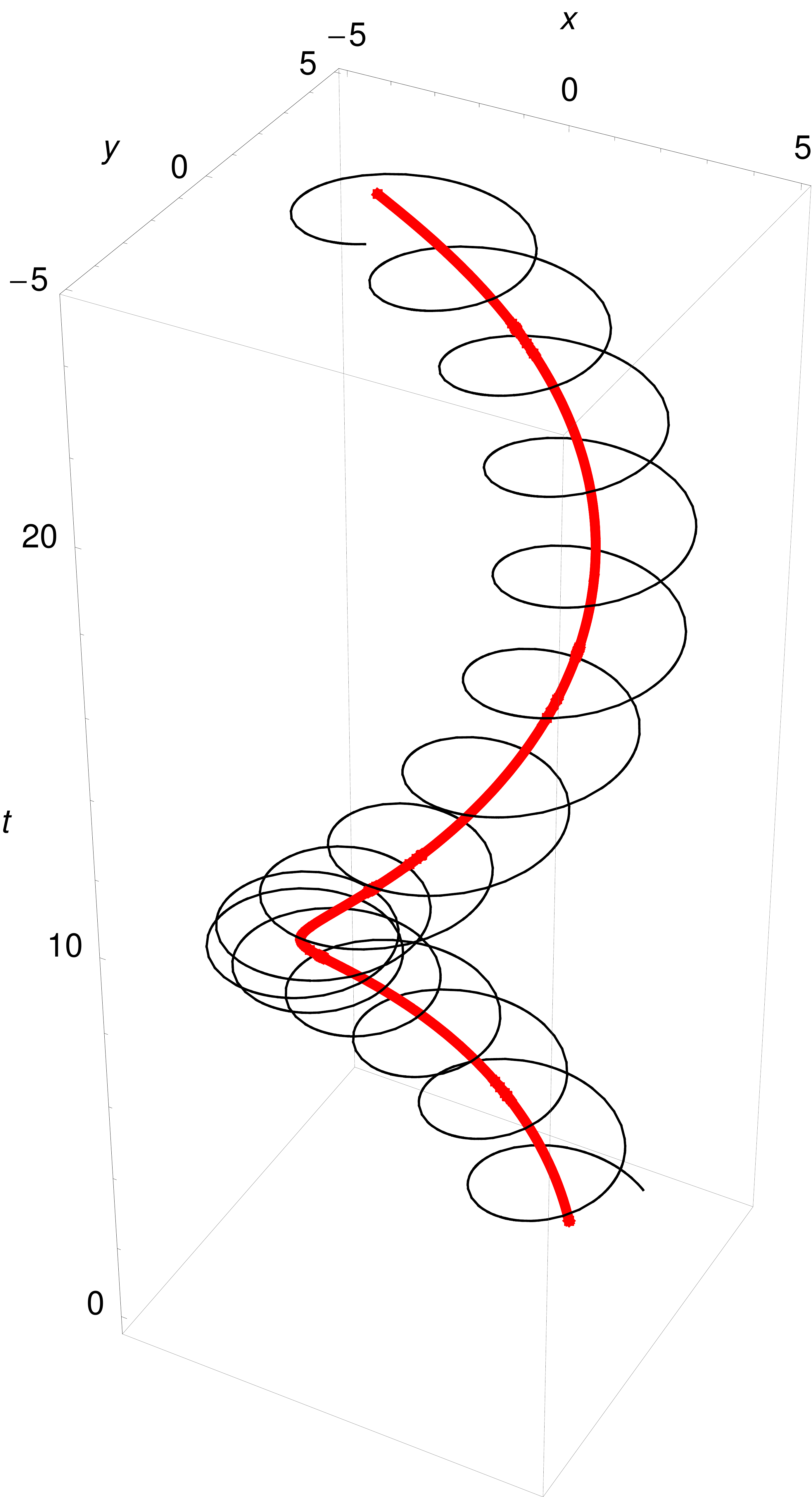}\;
\includegraphics[scale=.15]{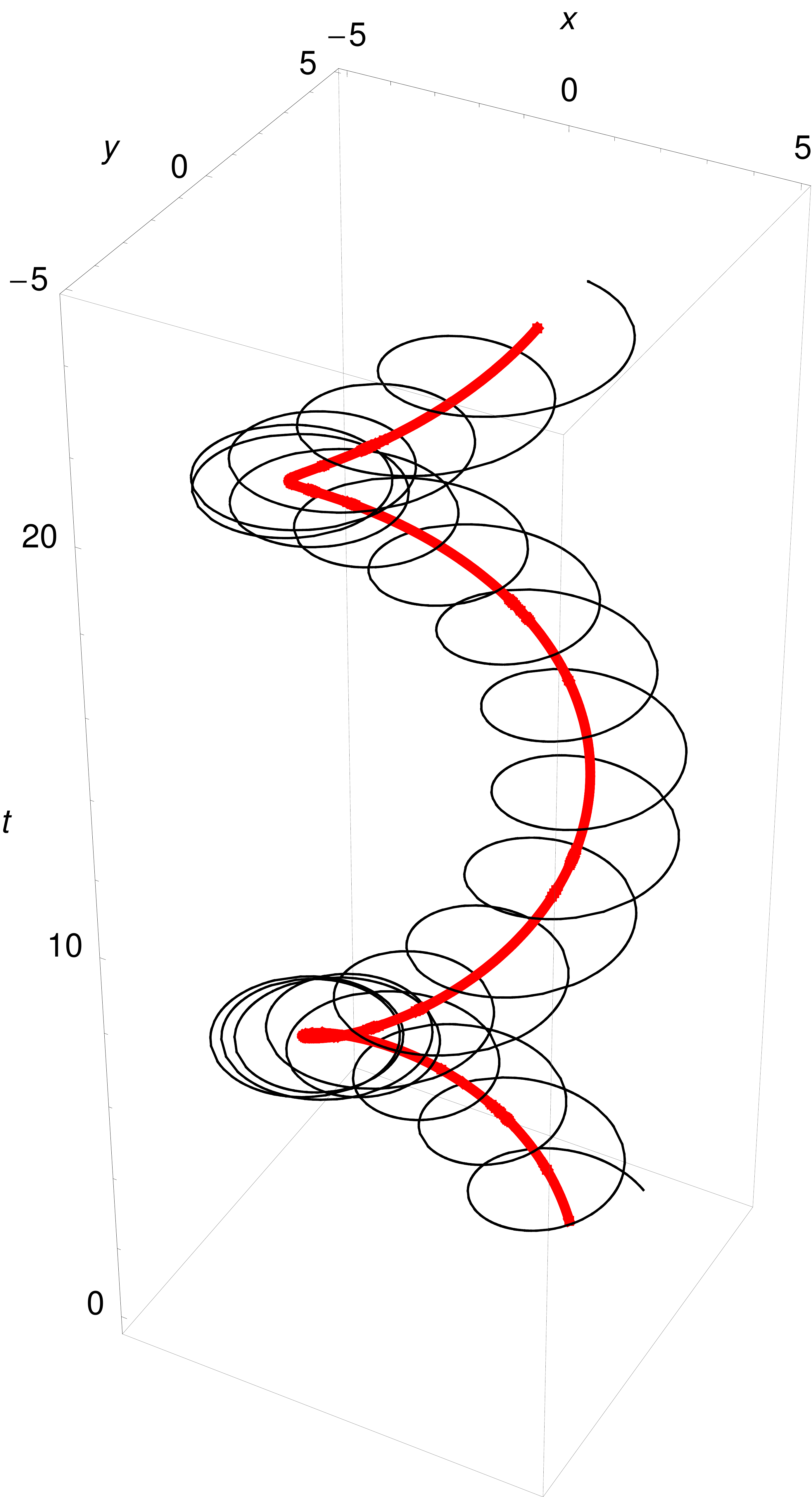}\;
\includegraphics[scale=.15]{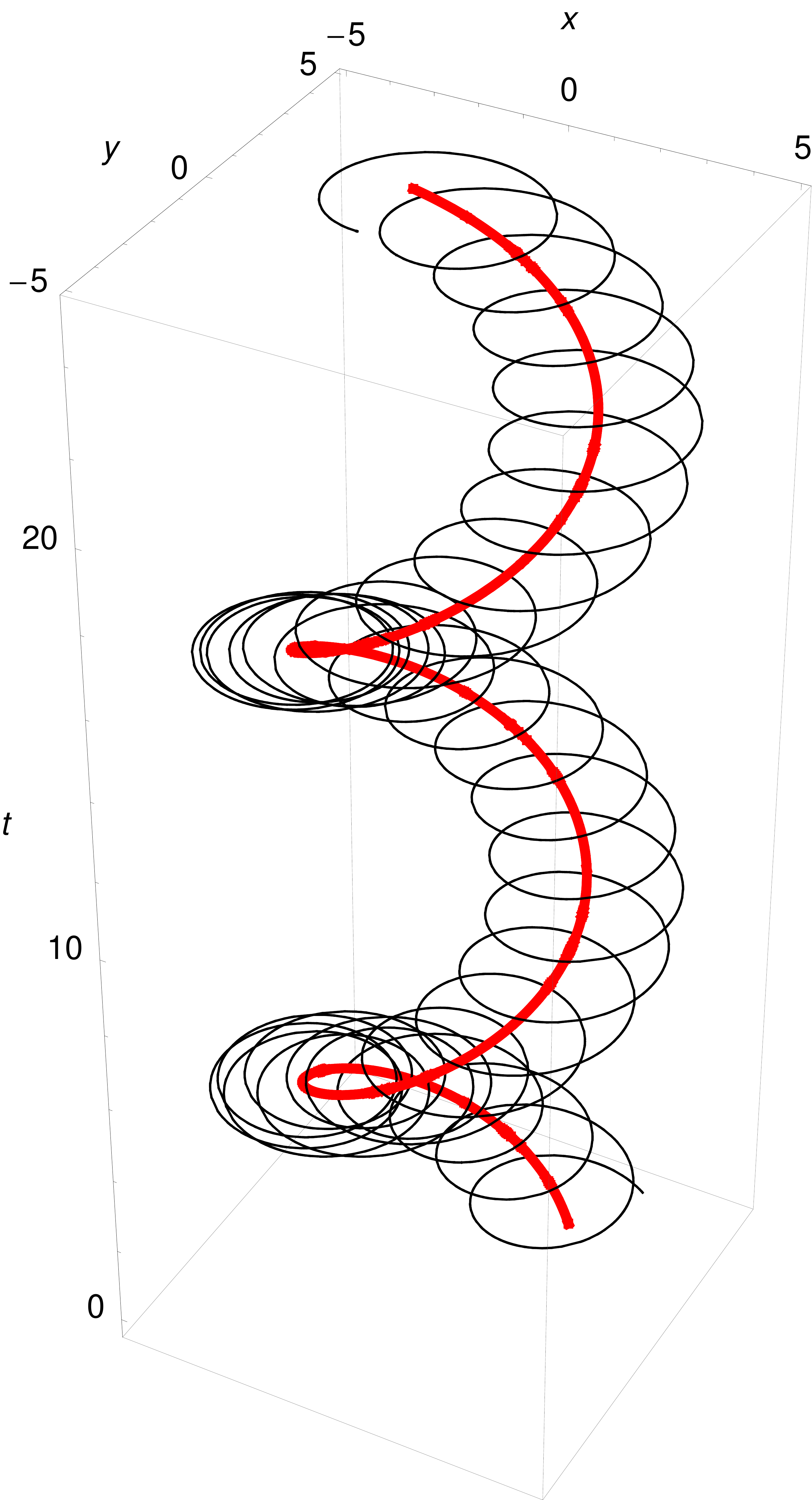}\\
\includegraphics[scale=.15]{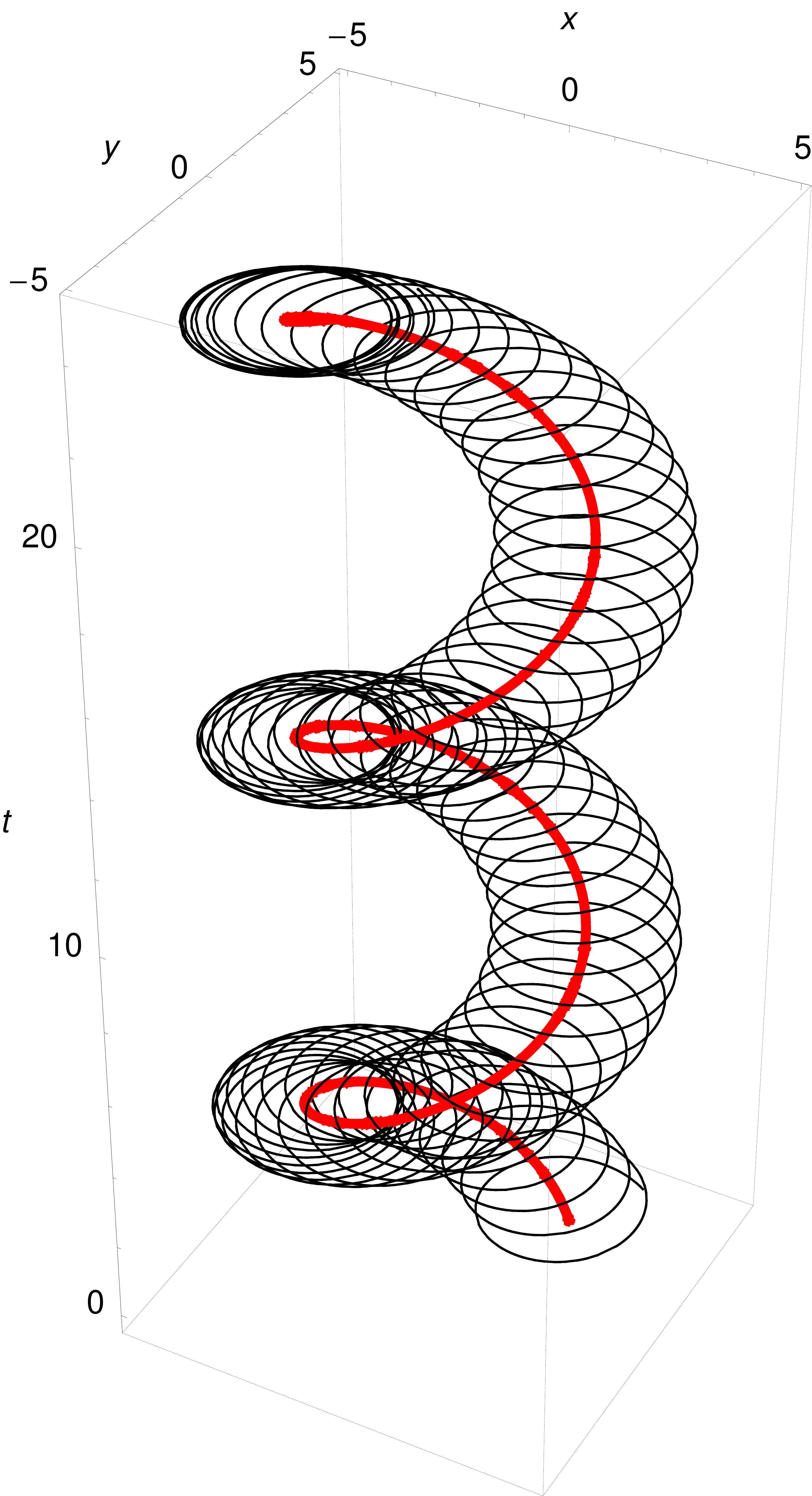}\;
\includegraphics[scale=.15]{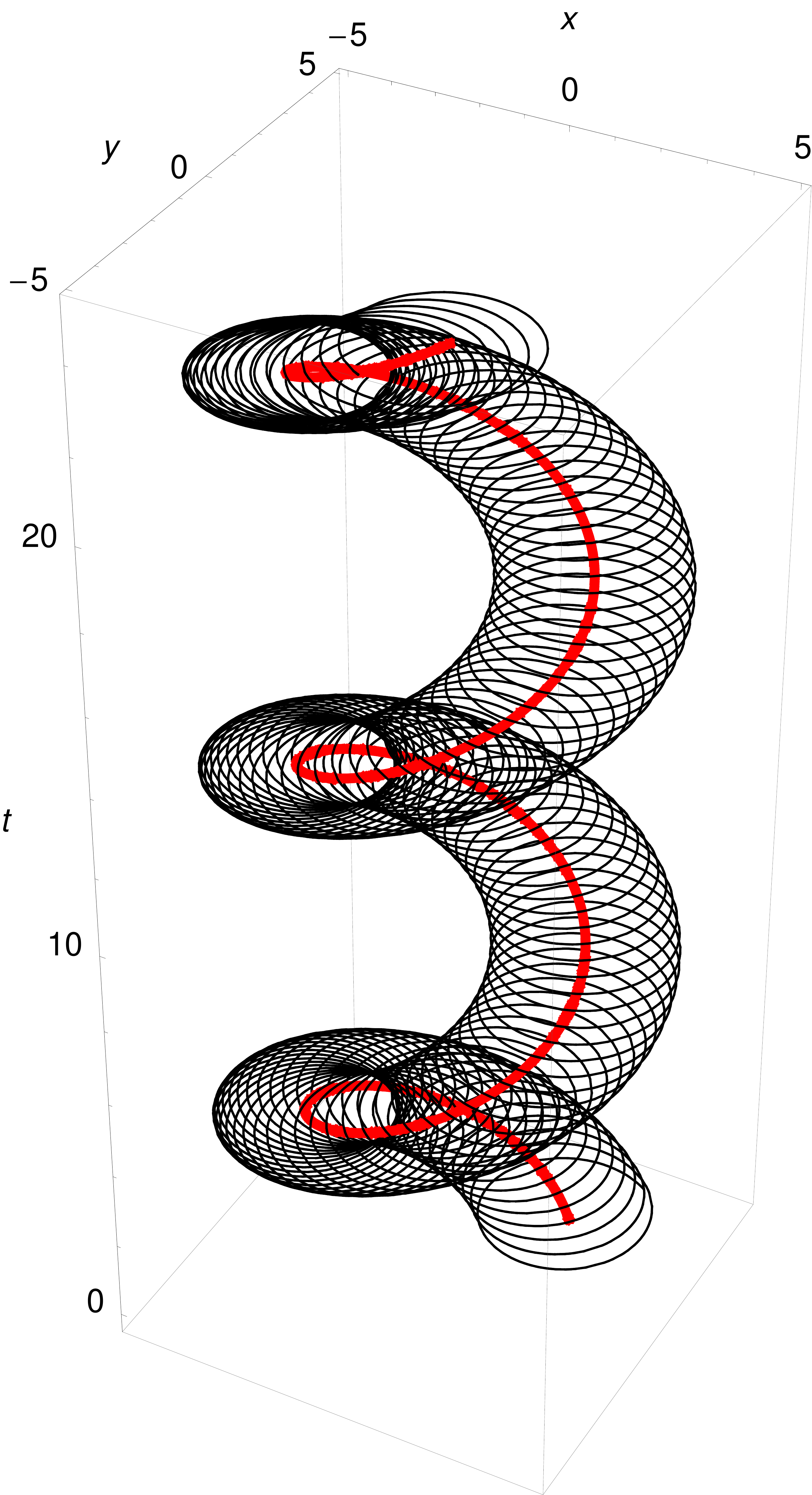}\;
\includegraphics[scale=.15]{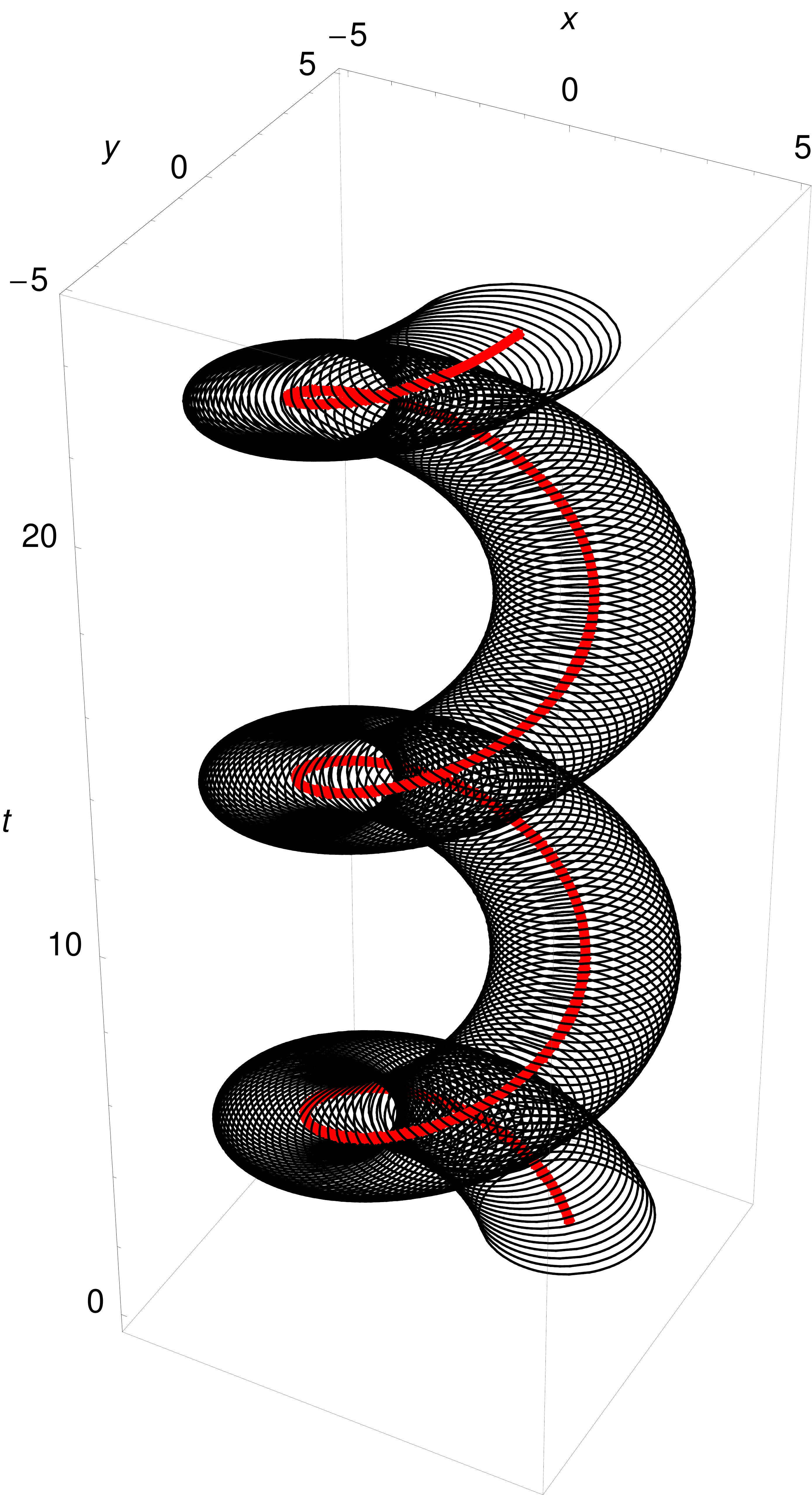}\\
\vspace{-7mm}
\end{center}
\caption{\it Trajectories unfolded into time with $eB\theta$ and the initial conditions kept constant
but the spring constant $k$ sweeping from very weak to very strong, $
 k= 0.5;\,1;\,2;\, 5;\, 10;\,20
 $. With increasing $k$ the ``whirling'' becomes stronger and stronger, whereas the guiding center motion changes slowly.
}
\label{BOFigure4}
\end{figure}

For $\mu_\pm\neq0$ the latter are
solved, putting $X_{\pm}=X_{\pm}^1+iX_{\pm}^2$, as
\beq
X_{\pm}(t)=e^{i\alpha_{\pm}t}X_{\pm}(0),
\quad\hbox{where}\quad
\alpha_{\pm}=\frac{mk+\omega_{\pm}^2}{m\mu_{\pm}}\,.
\label{pmOscmot}
\eeq
(In the purely magnetic case $k=0$, $\alpha_+=0$ and
$\alpha_-=-eB/m^*=-\omega^*$ as it should be, cf. \cite{DHexo,NCLandau}.)
Then $\vx(t)=\vX_+(t)+\vX_-(t)$ yields the motions,
some of which are depicted on  FIGs \ref{BOFigure1-b},
\ref{BOFigure4} and \ref{BO3D}, respectively.
Here, and in all subsequent figures, we choose $m=e=\theta=1$, and
also $k=1$ except in FIG. \ref{BOFigure4}. The [red] dotted line
indicates the guiding center.

Amusingly, all our trajectories are obtained by
combining uniform rotation following an ``epicycle'', whose center is rotating uniformly on a ``deferent'' circle --- as suggested  by Ptolemy of Alexandria in
the 1st century AD [but for planetary motion].
The  form of the trajectory depends on the
relative lengths [initial conditions] of
 $\vX_\pm(0)$ and on the frequencies, determined
by the relative strengths of the magnetic and
oscillator fields, respectively.

For the critical value
$B=B_c$ in (\ref{critB}),
\beqa
\omega_+^c=
-mk\theta,
\quad
\omega_-^c=
\displaystyle\frac{1}{\theta},
\qquad
\mu_+^c=
\frac{1}{\theta}(1+mk\theta^2)^2,
\quad
\mu_-^c=0.
\eeqa
The vanishing of $\mu_-$ implies that the $\vX_-$-oscillator drops out again from the symplectic form and the system
becomes singular \footnote{This can also be seen
from noting that det $[\Omega_{\alpha\beta}]=
\mu_+^2\,\mu_-^2$.}, while $\mu_+$ never vanishes,
since both the mass and the spring constant are supposed to be positive, $m,k>0$.
Then for $m^*\to0$ eqn. (\ref{pmOsceqmot}) can only be satisfied when
$\vX_-(t)=0$   for all $t$~: consistency requires
that the $\vX_-$-oscillator be switched off.
We are left, hence, with the $1$d $\vX_+$-oscillator alone.
The motions are therefore \emph{uniform rotations along circles about the origin},
\beq
x(t)=X_+(t)=e^{i\alpha^ct}X_+(0),
\qquad
\alpha^c=\alpha_+^c=\frac{k\theta}{1+mk\theta^2},
\label{BOHallmotion}
\eeq
which are in fact the Hall motions, cf. \cite{DHexo,NCLandau}. The Hall law requires indeed that the velocity be perpendicular to the (radial) force; the momentum is in turn
frozen into its ``Hall'' value in (\ref{Hallaw}).

This result
is consistent with the vanishing of $\vX_-(t)$ in (\ref{Ochircoord}). Let us also
stress that $\alpha^c$  in (\ref{BOHallmotion}) is the \emph{continuous limit} of  (\ref{pmOscmot})
when $B\to B_c$.

\section{Exotic Newton-Hooke symmetry}\label{NHsymmetry}

It has been recognized recently
\cite{AGKP,ZH-II}, that, off the critical case, $eB\theta\neq1$,  the NCLP admits the ``exotic'' two-fold central extension of the planar Newton-Hooke group as symmetry \cite{NewtonHooke}.
As hinted at in Ref. \cite{AGKP}, the symmetry  is readily recovered
using the chiral decomposition.

\subsection{Newton-Hooke symmetry in purely
magnetic background}

Let us first consider the purely magnetic case, $\vE=0$.

 The $\vX_+$-system is plainly invariant w.r.t.
translations,
\beq
\vX_+\to\vX_++\vc,
\eeq
for any constant vector $\vc$, and the associated conserved quantity, calculated using
\beq
\Omega_+(\delta\vX_+,\;\cdot\;)= -d\cP_+\,,
\eeq
cf. \cite{SSD}, is readily found, as
\beq
\cP^i_+=-eB\varepsilon^{ij}X_+^j\,.
\label{cP}
\eeq
 Its commutation relation,
calculated using
\beq
\{\cP_+^i,\cP_+^j\}=-\Omega(\delta X_+^i,\delta X_+^j)
\eeq
\cite{SSD} is, moreover,
\beq
\{\cP_+^i,\cP_+^j\}=-eB\varepsilon^{ij},
\label{PPcom}
\eeq
i.e., that of the
\emph{Heisenberg group} with central extension parameter $(-eB)$.

Extending the $\vX_+$-only translations onto the whole phase space trivially, i.e., by $\vX_-\to\vX_-$ [so that $\cP^i_-=0$], we get
ordinary translations  $x^i\to x^i+c^i$, and our total conserved quantity, expressed in the original coordinates, reads
\beq
{\bm\cP}={\bm\cP}_++{\bm\cP}_-,
\quad\hbox{where}\quad
\cP^i=p^i-eB\varepsilon^{ij}x^j,
\eeq
as found before in \cite{ZH-II}. Its commutation relations are still given in (\ref{PPcom}).

 Similarly, the $\vX_-$-system has Hamiltonian structure
 $\Omega_-$ and $H_-$ which is that of a $1$d harmonic oscillator,  and for which
\beq
\vX_-\to\vX_-+ R(-\omega^*t)\va
\label{X-rot}
\eeq
is  a symmetry for any planar vector $\va$. The associated conserved quantity,
\beq
\vcK_-=m(1-eB\theta)^2\,R(\omega^* t)\dot{\vX}_-,
\label{vcK}
 \eeq
satisfies
\beq
\{\cK^i_-,\cK^j_-\}=(1-eB\theta)eB\,\varepsilon^{ij},
\label{KKcomm}
\eeq
i.e., the Heisenberg commutation relation but with ``exotic'' central parameter $(1-eB\theta)eB$.
Extending the action (\ref{X-rot}) trivially to the $\vX_+$-part provides us with
 ``imported boost'' symmetry $\vx\to\vx+R(-\omega^*t)\va$ \cite{NewtonHooke,ZH-II},
 and with [total] conserved quantity
\beq
\vcK=(1-eB\theta)\,R(\omega^* t)\,\vp.
\eeq

Planar rotations by angle $\varphi$,  acting diagonally \cite{AGKP} i.e. as $\vX_{\pm}\to R(\varphi)\vX_{\pm}$,
leave both dynamics invariant and provide us with conserved angular momenta, namely
\beq
\cJ_+=\frac{eB}{2}\vX^2_+,
\qquad
\cJ_-=-(1-eB\theta)\frac{eB}{2}\vX^2_-\,.
\label{pmangmom}
\eeq
Expressed in terms of the original coordinates,
\beq
\cJ=\cJ_++\cJ_-=\vx\times{\vp}+\frac{eB}{2}\vx^2+
\frac{\theta}{2}\,\vp^2,
\eeq
cf. \cite{DHexo,ZH-II}.
We mention, for completeness, that the Hamiltonian
$H={\vp^2}/{2}$  is the conserved quantity
associated with time translation symmetry $t\to t-\tau$ acting on the  respective chiral oscillator phase spaces.

\subsection{Newton-Hooke symmetry in magnetic +
harmonic trap background}


According to recent results, the
exotic Newton-Hooke symmetry of the purely-magnetic
problem can be extended to a constant electric
field \cite{ZH-II}. But in Sect.
\ref{Chiral} A we have shown that the chiral decomposition remains valid even
after switching on
a constant electric field. It should  hence be possible, in principle,
to deduce its symmetries from those of the chiral
components. For the $\vX_+$ component this would be easy; in particular, $\vX_+$-alone-translations
are plainly symmetries~: they merely shift the
guiding center of the motion.
The $\vX_-$-alone-symmetries are, however, a bit more
complicated, since the Hamiltonian $H_-$ in
(\ref{EsplitHam}) corresponds to a forced $1$d oscillator, and the construction of
\cite{AGKP} would therefore require  extension.
Considering the construction in \cite{ZH-II}
satisfactory, we do not pursue this issue here.

The   Landau problem with a confining harmonic trap does,
however, enter into the framework
of Alvarez et al.  \cite{AGKP}, as the latter  applies to any system which splits into two
chiral oscillators.
Spelling out their general formulas, we present here some details on the exotic
 Newton-Hook symmetry of the problem.

Firstly, the oscillator potential breaks the translation-invariance; both chiral oscillators admit instead independent ``imported boost'' symmetries, namely
\beq
\vcK_{\pm}=m\mu_{\pm}^2\,R(-\alpha_{\pm}t)\dot{\vX}_{\pm},
\eeq
with commutation relations
\beq
\{\cK_{\pm}^i,\cK_{\pm}^j\}=-\mu_{\pm}\,\epsilon^{ij},
\qquad
\{\vcK_{+},\vcK_{-}\}=0.
\eeq
The total angular momentum and energy,
\beqa
\cJ&=&\cJ_++\cJ_-,\qquad\;
\cJ_{\pm}=\frac{1}{2}\mu_{\pm}\,\vX_{\pm}^2\,,
\\[6pt]
H&=&H_++H_-,\qquad
H_{\pm}=
\frac{mk+\omega_{\pm}^2}{2m}\,
\vX_{\pm}^2,
\eeqa
respectively, are associated to the diagonal action of rotations and time translations
on the respective oscillator phase spaces. Note that $H_\pm=\alpha_{\pm}\cJ_{\pm}$.

\section{Quantization}

So far, all our investigations have been purely classical. It is
not difficult to quantize our system, though, along the same lines as in the ordinary case \cite{NCLandau,NewtonHooke}. For simplicity, we only consider  the purely magnetic
$\vE=0$.

Classical quantities are promoted to operators [distinguished by ``hats'']; Poisson brackets
are replaced by
$i/\hbar$-times commutators.

Assuming first that we are in the non-critical regime, $eB\theta\neq1$, we
observe that the quantum  Hamiltonian is
expressed in terms of the conserved quantity $\vcK$ alone: putting
\beq
\hat{a}=\hat{\cK}^1+i\hat{\cK}^2
\quad\hbox{and}\quad
\hat{a}^\dagger=\hat{\cK}^1-i\hat{\cK}^2,
\eeq
the Hamiltonian reads indeed
\beq
\hat{H}=\hat{H}_-=
\frac{\hat{a}^\dagger\hat{a}}{2m(1-eB\theta)^2}+\hbar\frac{\omega^*}{2}.
\label{QHam}
\eeq

Note that $\hat{a}^\dagger$ and $\hat{a}$
are ``$\vX_-$-only operators''by (\ref{vcK}).
Owing to (\ref{KKcomm}) we have,
\beq
[\hat{a},\hat{a}^\dagger]=2\hbar(1-eB\theta)m\omega.
\label{aadagcom}
\eeq
$\hat{a}$ and $\hat{a}^\dagger$
are annihilation and creation operators acting
on Fock space, $\hat{a}|0\rangle=0$, so that
$|n\rangle = (\hat{a}^\dagger)^n|0\rangle,\, n=0,\,1,\,\dots$ is an eigenvector
of $\hat{a}^\dagger\hat{a}$ with eigenvalue
$2\hbar(1-eB\theta)m\omega\, n$.
It follows that the energy levels are
\beq
E_n=\hbar \omega^*\big(n+\frac{1}{2}\big)=
\hbar\frac{eB}{m^*}\big(n+\frac{1}{2}\big),
\quad n=0,1,\dots,
\label{CMpectrum}
\eeq
consistently with \cite{NCLandau}.  Note that the energy only depends on the $\vX_-$-dynamics
\footnote{The degeneracy of the energy levels is plainly lifted when an electric field is added.}, its sign  is that of
$\mu_-=-eB(1-eB\theta)$ and flips therefore when the critical value is crossed.

In the same spirit, consider the oscillator representation of
the other conserved quantity, \textit{viz.} the magnetic momentum $\bm{\cP}$,
\beq
\hat{b}=\hat{\cP}^1+i\hat{\cP}^2,
\quad\hbox{and}\quad
\hat{b}^\dagger=\hat{\cP}^1-i\hat{\cP}^2.
\eeq
These are, by (\ref{cP}), ``$\vX_+$-only operators'' which satisfy
\beq
[\hat{b},\hat{b}^\dagger]=2i\hbar m\omega.
\label{bbdagcom}
\eeq
$\hat{b}$ and $\hat{b}^\dagger$
are therefore again annihilation and creation operators, so that
$|p\rangle  = (\hat{b}^\dagger)^p|0\rangle $ is an eigenvector
of $\hat{b}^\dagger\hat{b}$ with eigenvalues
$2\hbar eB\,p$ with $p=0,1,2,\dots$.
Our new oscillator-operators do not intervene into
the energy, but they do enter into the
\emph{total angular momentum operator},
\beq
\hat{J}=\frac{1}{2eB}\left(\hat{b}^\dagger\hat{b}
-\frac{1}{1-eB\theta}\,\hat{a}^\dagger\hat{a}
\right),
\eeq
which has, therefore, eigenvalues
 \cite{AGKP}
\beq
J_{p,n}=\,\hbar\,\Big({\rm sg}(\mu_+)p-{\rm sg}(\mu_-)n
+\frac{1}{2}({\rm sg}(\mu_+)-{\rm sg}(\mu_-)\Big),
\qquad p,n=0,1,2,\dots\,
\eeq
Thus, the spectrum of the angular momentum is unbounded in the ``subcritical regime'' $eB\theta<1$ and becomes bounded from one side when
we cross over into the ``supercritical regime'' $eB\theta>1$
cf. \cite{AGKP,NCLandau}.

Our quantum Hilbert space is thus composed of wave functions
\beq
\psi=|p,n\rangle=|p\rangle\,|n\rangle.
\eeq

The quantization in the critical case $B=B_c$ is postponed to  the next Section.

Skipping details, we would like to mention that the Landau problem with
confining oscillator symmetry could be quantized along the same line, namely by replacing the
magnetic translational
$\bm{\cP}$-symmetry by the ``imported boost symmetry'', $\vcK_+$.

\section{Study of the ``phase transition''
$m^*\to0$}\label{critical}

\subsection{Classical aspects}

It is amusing to study the ``phase transition''
$m^*\to0$ in some detail. Let us start with the classical mechanics. A look at our plots reveals that the trajectories seem, somewhat surprisingly, to keep their body shapes as $m^*\to0$ --- although we know that most of them
become inconsistent. How can this come about~?
A simple explanation is as follows: write the
velocity-momentum relation [the first equation in
(\ref{exoeqmot}), or its $X_-$-counterpart in the chiral decomposition, eqn.
(\ref{chireqmot}) or in (\ref{pmOsceqmot})]
in the form
\beq
\mu_-\,dX_-^i=\Big(\,\dots\,\Big)dt.
\eeq
Then, as $\mu_-\to0$, we have two alternatives:

\begin{enumerate}
\item
either $\big(\,\dots\,\big)=0$, which eliminates all initial conditions \emph{except those which are consistent with the Hall law};

\item
or \dots the ``motion'' should be \emph{instantaneous},
$dt=0$~!
\end{enumerate}

An intuitive understanding of the transition from the
``subcritical'' to  ``supercritical'' regime, and of
the appearance  of our strange ``instantaneous motions'' in particular, is that, being proportional to $(\mu_-)^{-1}$,
 the frequency of the
$\vX_-$-oscillator goes to infinity as
$B\to B_c$. This implies that the body of the trajectory is unchanged
 but it is described more and more rapidly. When $\mu_-=0$ we have the singular
case, and only the Hall motions keep on advancing in time at a regular [namely the Hall] velocity, whereas all other motions are ``flattened''
to $t=\const$. Changing the sign of
$\mu_-$ amounts, furthermore, to changing the orientation of the trajectories~: our
$\vX_-$-oscillator flips over its chirality.
All this is confirmed by the plots in FIG.s \ref{pmagsweep}--\ref{BE3Dsweep}--\ref{BO3D} below.
FIG. \ref{BO3D} provides, in the oscillator case, particular
insight~: when the critical value is approached, the
``non-Hall'' trajectories speed up since the
frequencies, inversely proportional to ${\mu_-}$, diverge.
But the frequency of the central line representing the guiding center, which performs Hall motion,
remains finite, namely as given in (\ref{BOHallmotion}).

Another interesting aspect is that when the critical value $B_c$ is crossed, all commutation relations in (\ref{exoPB}), change sign.
\begin{figure}
\begin{center}
\includegraphics[scale=.2]{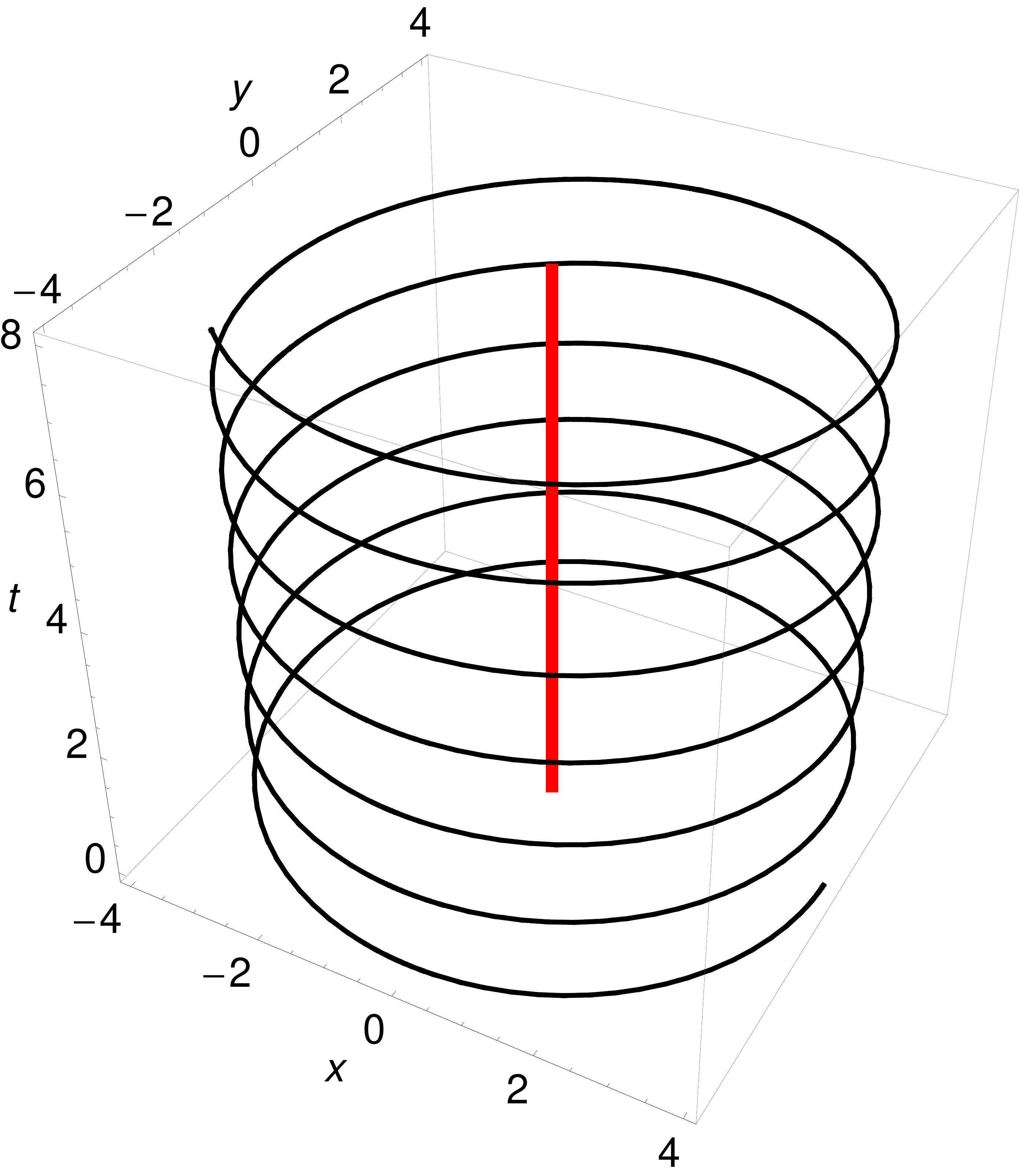}\quad
\includegraphics[scale=.2]{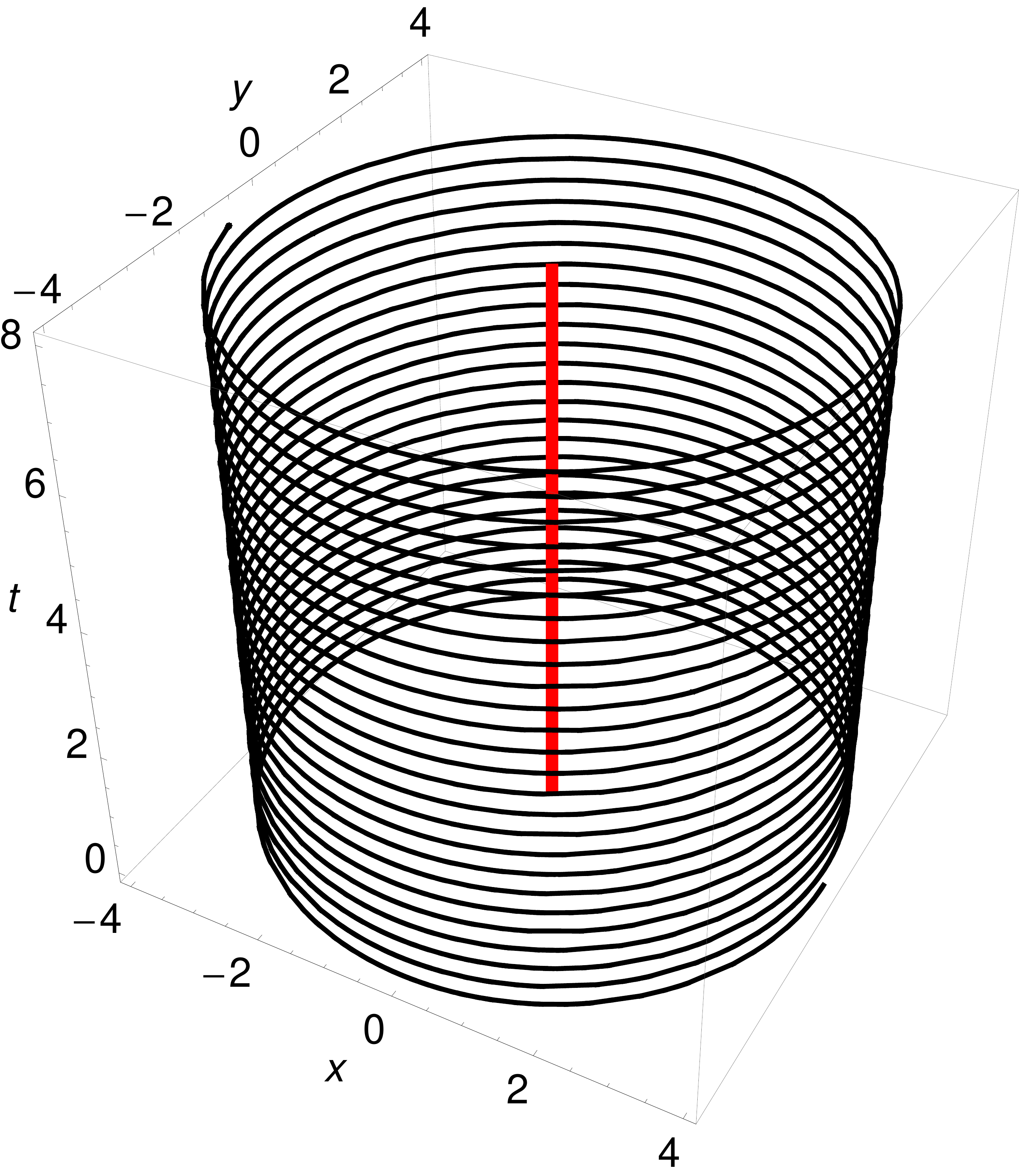}\\
\includegraphics[scale=.2]{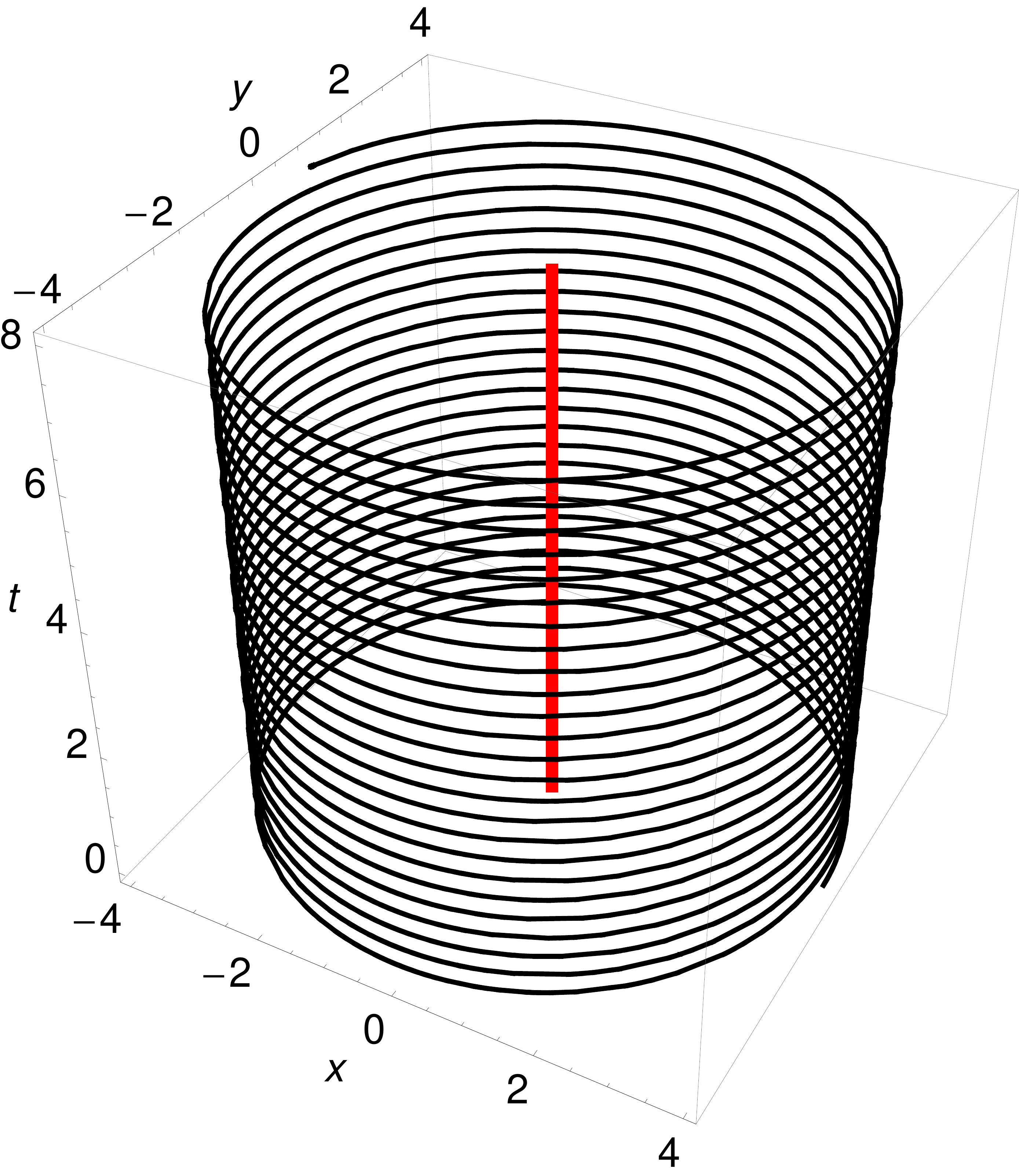}\quad
\includegraphics[scale=.2]{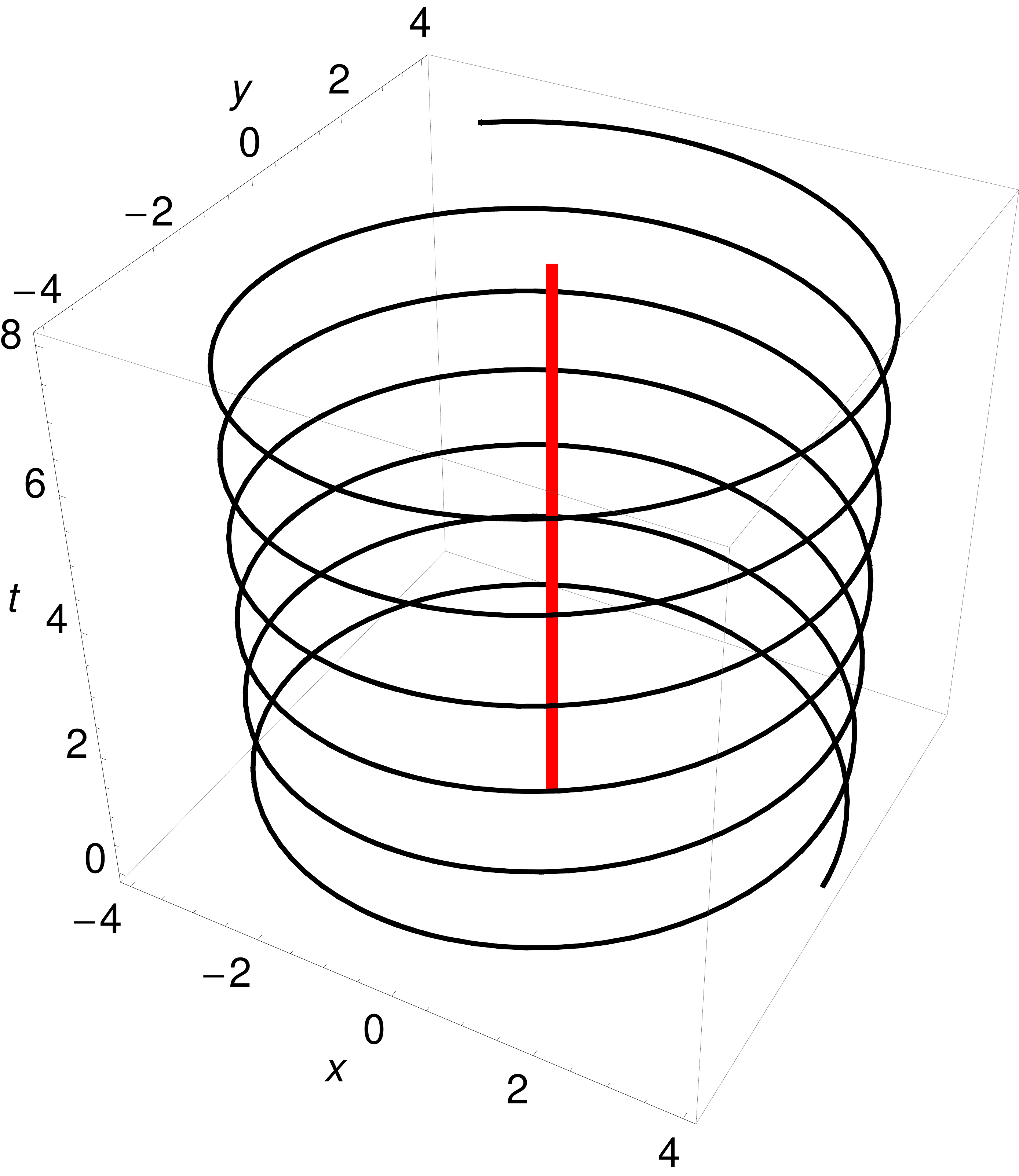}
\vspace{-7mm}
\end{center}
\caption{\it Trajectories in the purely-magnetic problem
for $eB\theta=0.8,\,0.95,\,1.05,\,1.20$.
When $B\theta<1$ is increased starting from the weak-field regime, the trajectory ``speeds up'' and becomes hence `` denser'' until it reaches the critical regime.
For $eB\theta=1$, all trajectories become ``instantaneous'' --- with the exception of the
``Hall ones'', which are now simple points.
After crossing the critical value, the chirality is reversed
and the curves
change orientation.
This also explains why the angular momentum
 becomes bounded from one side after crossing the
critical value.}
\label{pmagsweep}
\end{figure}

\begin{figure}
\begin{center}
\includegraphics[scale=.22]{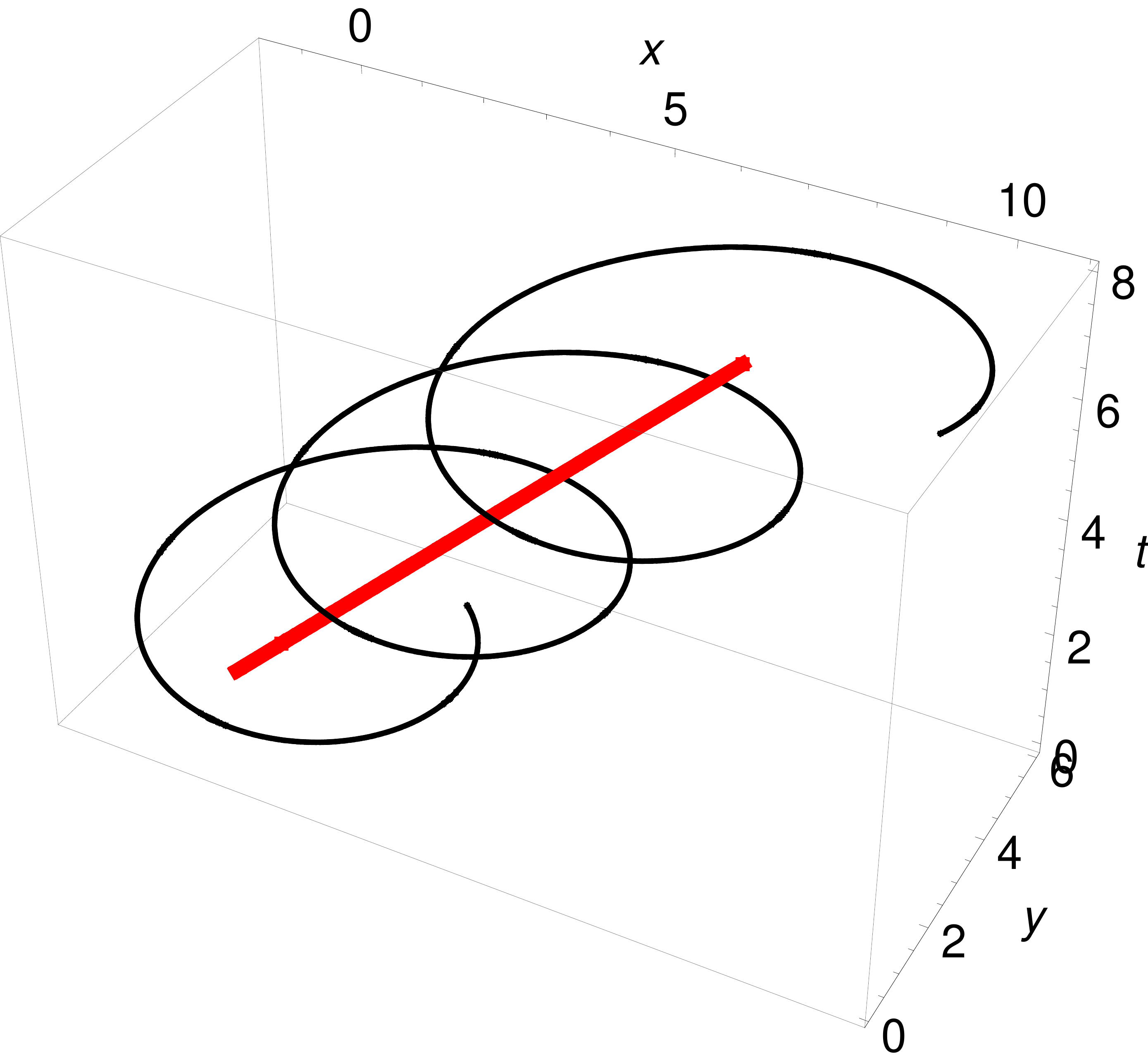}\quad
\includegraphics[scale=.22]{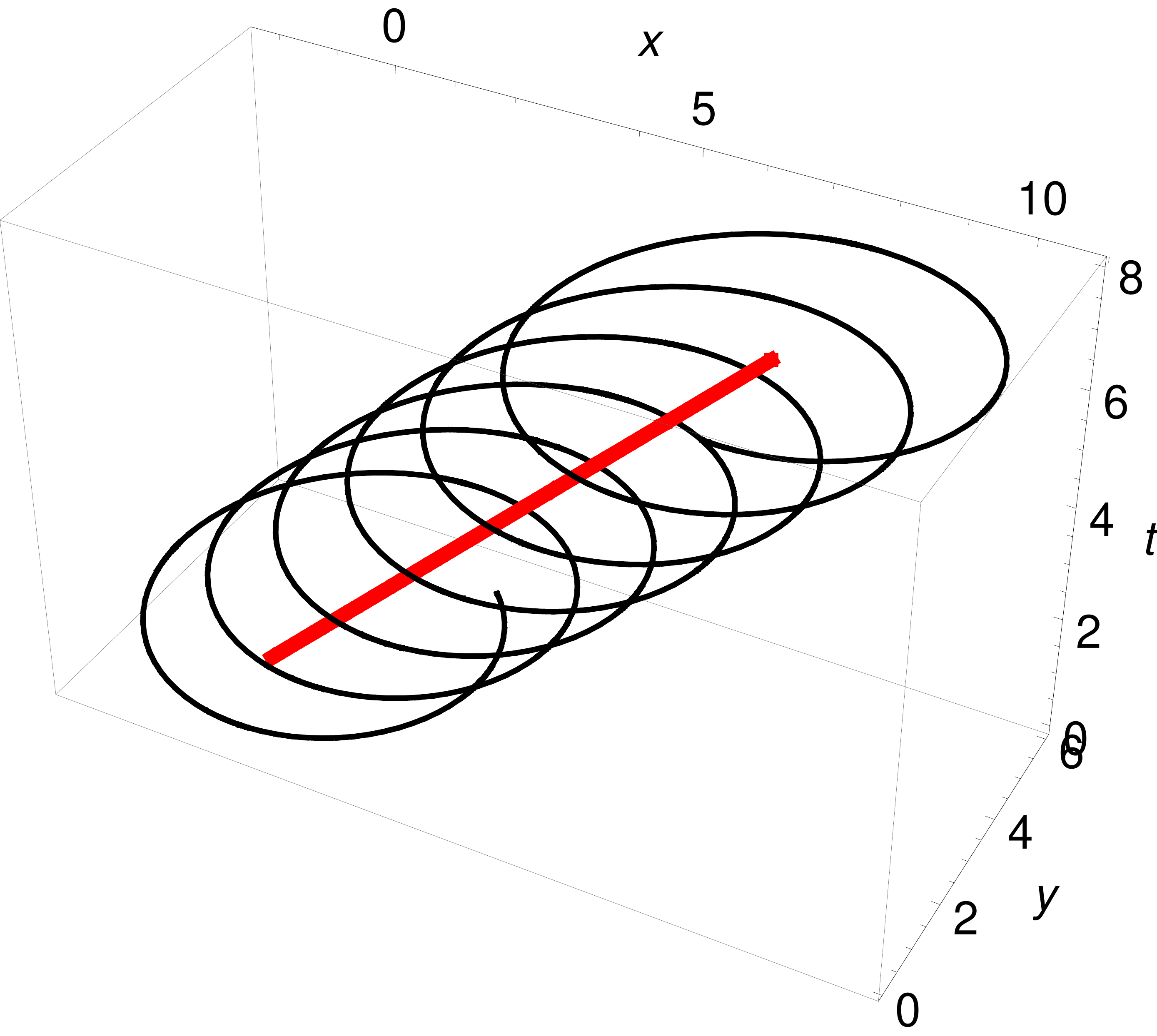}\\
\includegraphics[scale=.22]{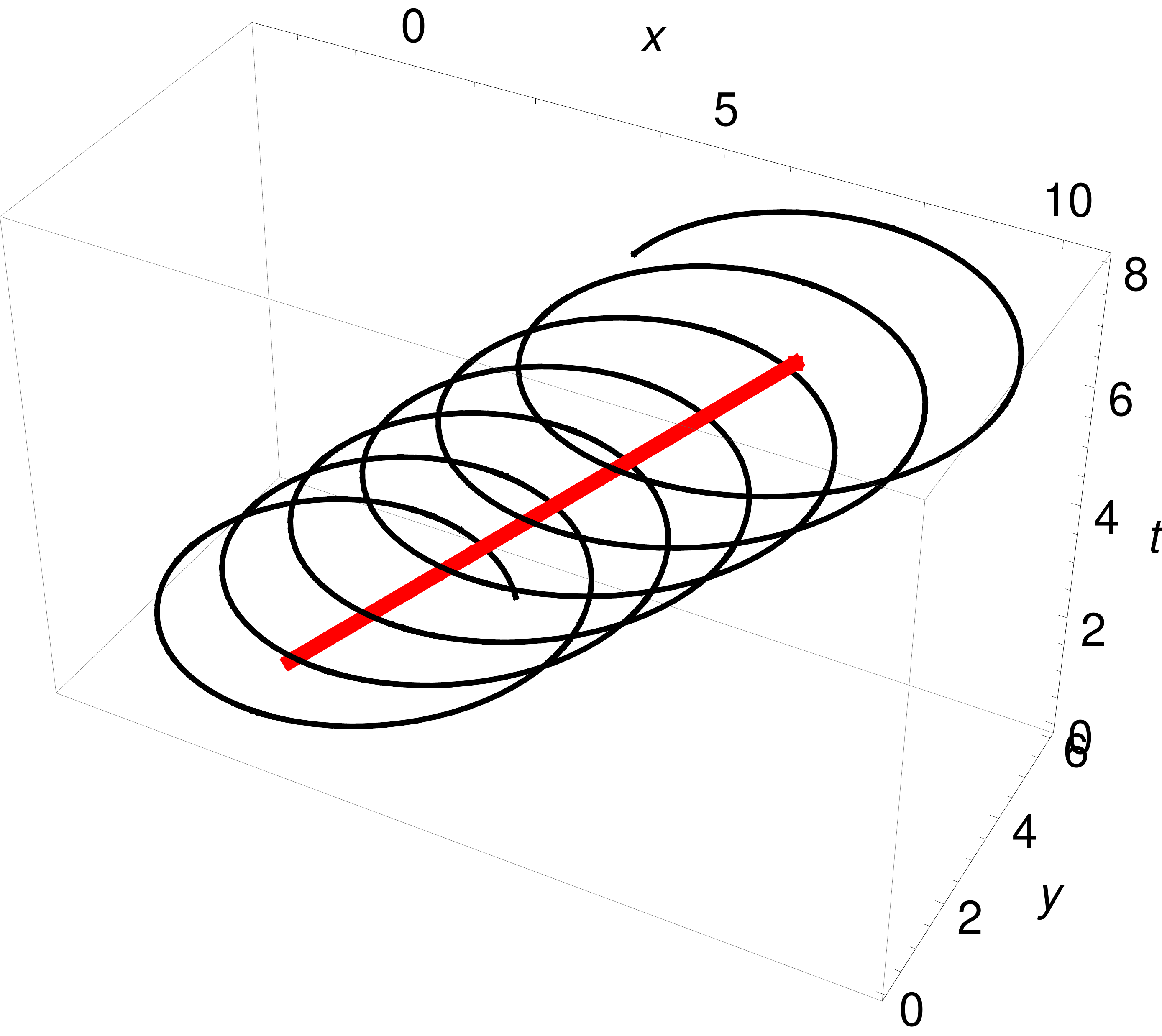}\quad
\includegraphics[scale=.22]{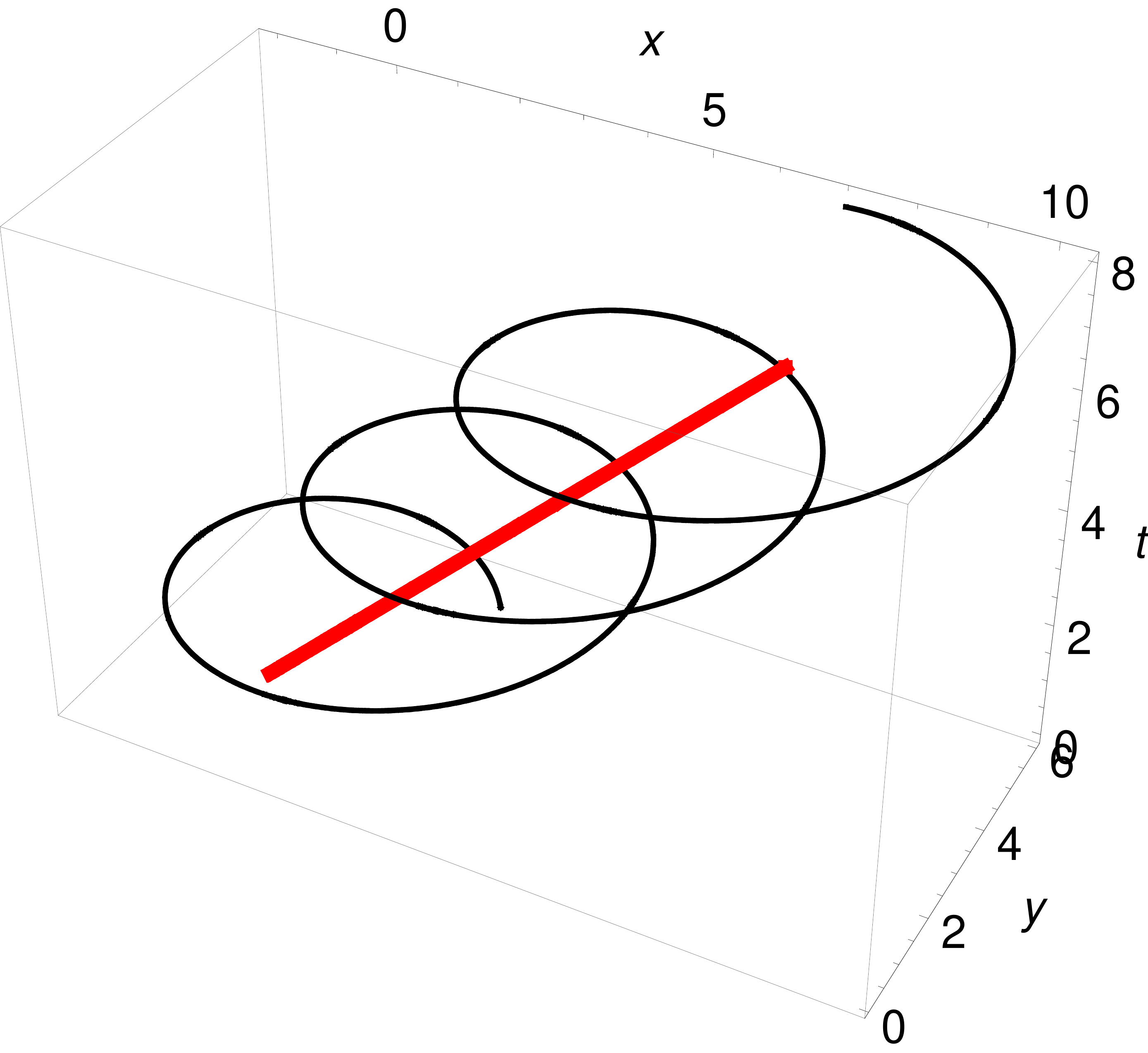}
\vspace{-7mm}
\end{center}
\caption{\it Trajectories in the magnetic plus constant electric field problem unfolded into time,
with the electric field
oriented into the second direction
and $eB\theta$ sweeping through
the values $0.6,\,0.8,\,1.20,\, 1.4$.
 The behavior is similar to the one in the purely magnetic problem,
except for the Hall drift of the guiding center.
}
\label{BE3Dsweep}
\end{figure}

\begin{figure}
\begin{center}
\includegraphics[scale=.19]{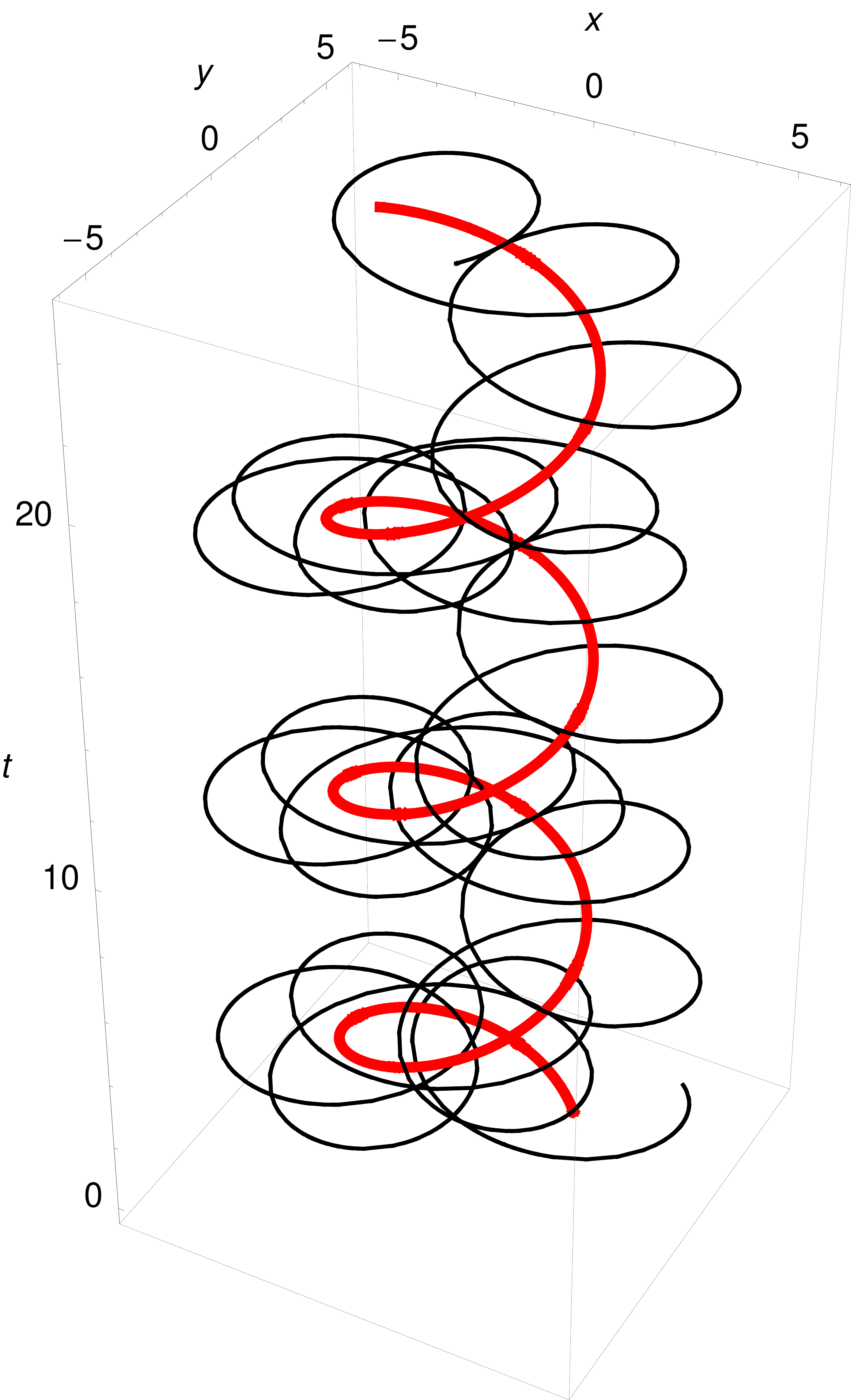}\quad
\includegraphics[scale=.19]{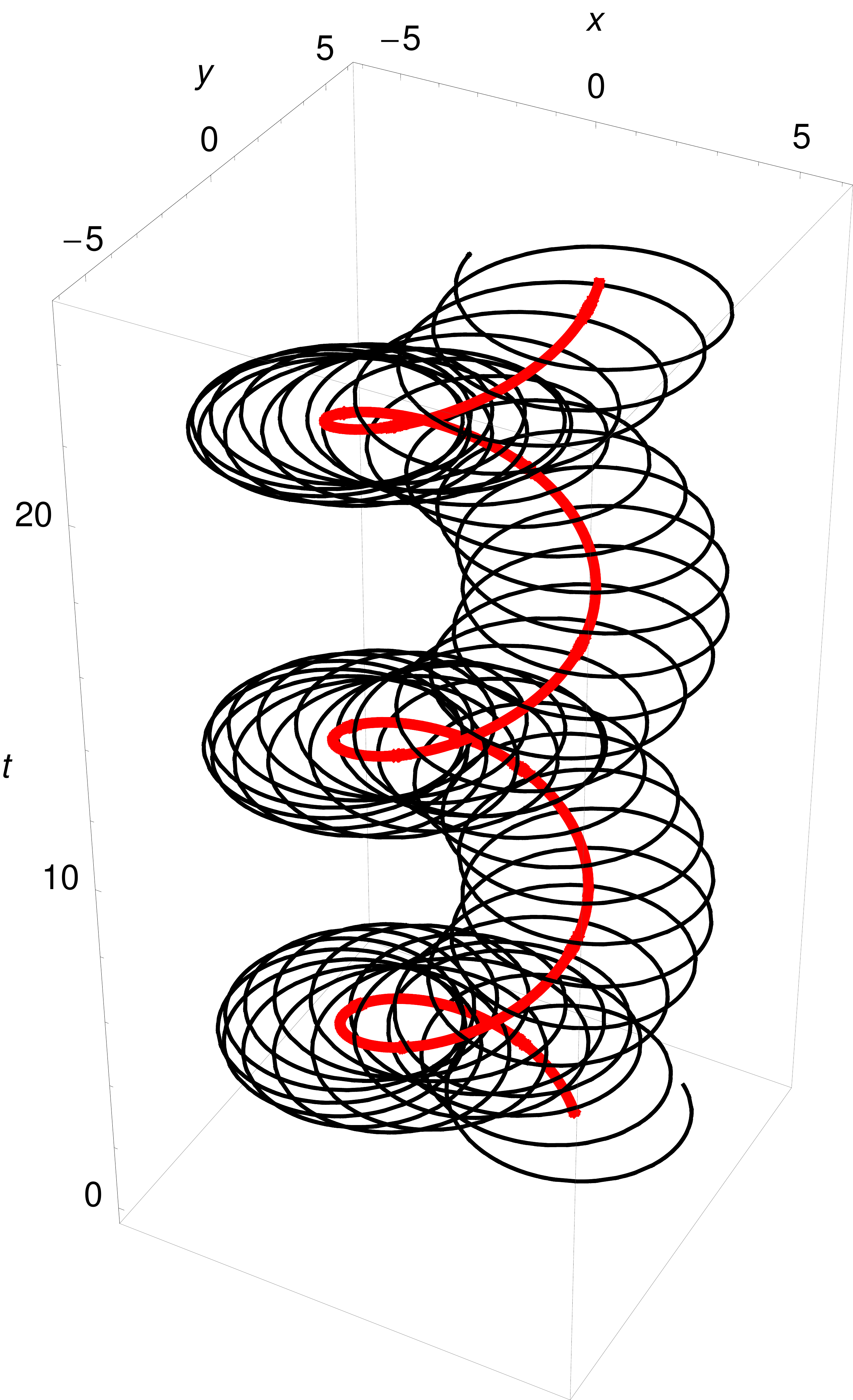}\\
\includegraphics[scale=.19]{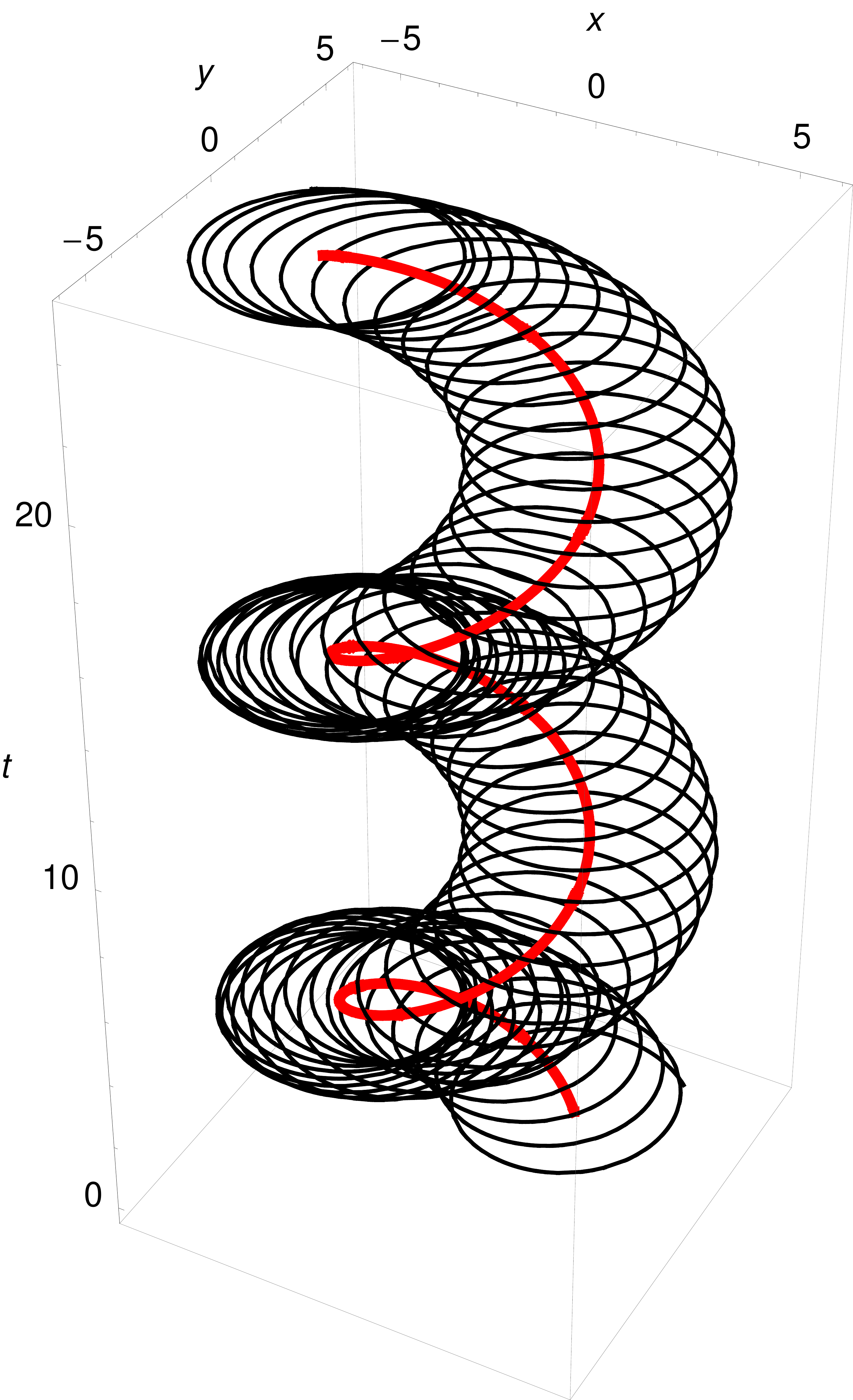}\quad
\includegraphics[scale=.19]{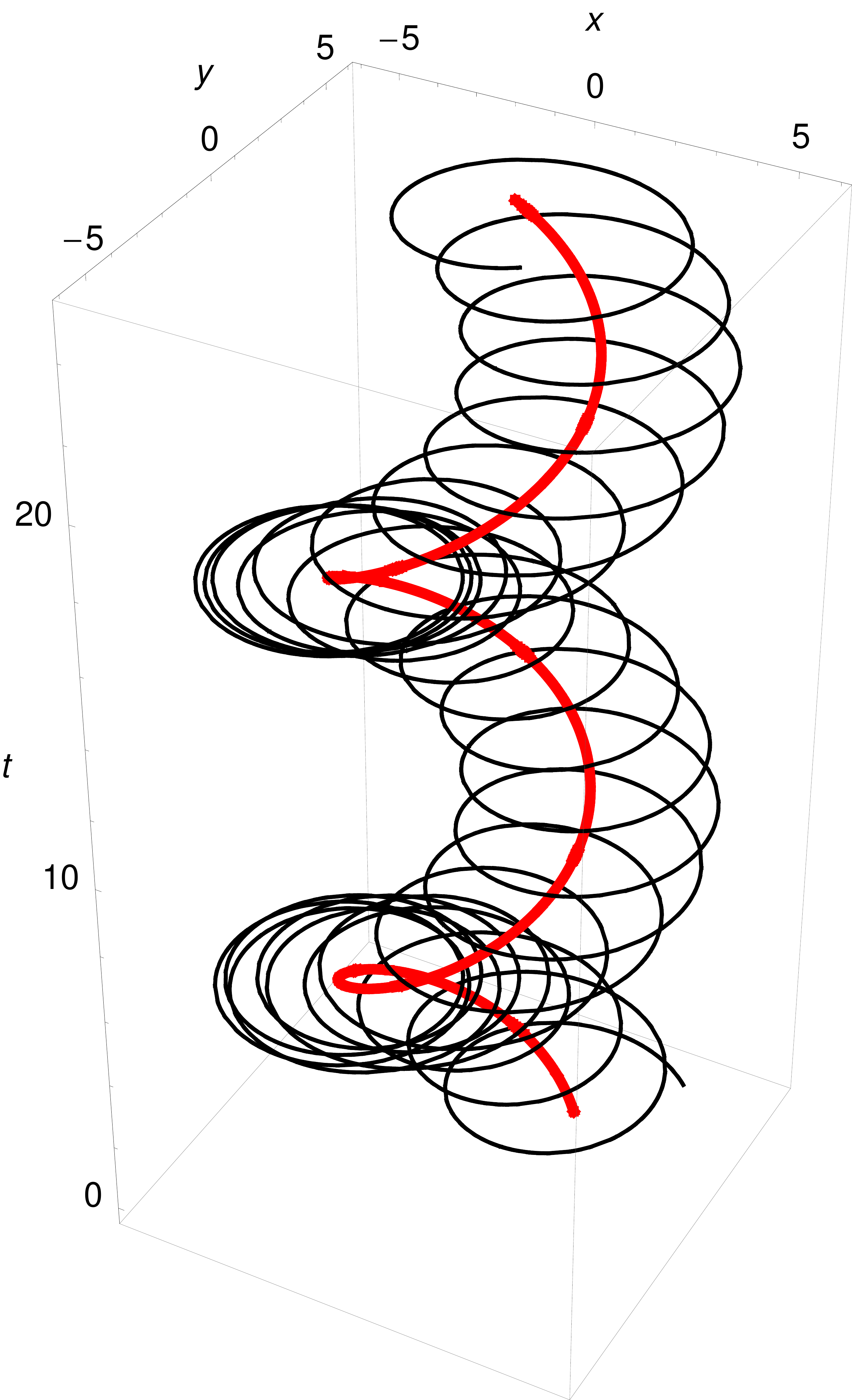}
\vspace{-7mm}
\end{center}
\caption{\it Trajectories in the magnetic- plus harmonic trap problem unfolded into time, for
$k=2$ and
$eB\theta=0.5,\,0.8,\,1.20,\, 1.5$.
The asymmetry between the sub- and supercritical regimes is caused by the harmonic background. When the critical value is approached,
``non-Hall'' trajectories speed up
but the frequency of the central line indicating the guiding center
remains finite.
}
\label{BO3D}
\end{figure}

\subsection{Quantization in the critical phase}

Turning to  quantization in the critical case,
 we observe that, when $m^*\to0$,
only those quantum states can
survive which are killed by $\hat{a}^\dagger\hat{a}$,
due to the $(1-eB\theta)^{-2}$ factor
in the Hamiltonian (\ref{QHam}).
Such states belong therefore to a subspace of the original
Hilbert space, --- namely that of the \emph{lowest Landau level}, characterized by
$n=0$ in (\ref{CMpectrum}). Although the ground-state energy $E_0$ diverges as $m^*\to0$, the ``good'' [meaning "Hall"] states themselves \emph{do} survive the transition: they belong to the Fock space of $\hat{b}$ and $\hat{b}^\dagger$
alone, $|p\rangle=(\hat{b}^\dagger)^p|0\rangle$, where $\hat{b}|0\rangle=0$.

The reduced Hilbert space can also be represented
by the ``Bargmann-Fock'' wave functions of the form
\beq
\psi(z)=f(z)e^{-\vert z\vert^2/4\theta},
\label{redwf}
\eeq
 with $f(z)$
analytic in
$ z=(X_+^1+iX_+^2)$
\cite{DHexo,NCLandau}. Physicists have long appreciated this point \cite{QHE,GJ,Dunne}.

It is the Hamiltonian operator $\hat{H}$ which changes when $B\to B_c$, namely by renormalizing $\hat{H}$ by subtracting the
ground-state term $\hbar\omega^*/2$.
It becomes, hence,
\emph{identically zero} in the purely magnetic case,
or the potential alone, $V$, otherwise.  Moreover,
$\vX_-$ is frozen and the
dynamical degree of freedom represented by it is lost,
so that  our reduced ``Hamiltonian'' [i.e. the
potential] is a function of $\vX_+$ alone.  The latter
has, however \emph{non-commuting coordinates}.

This is the  ``Peierls substitution'' \cite{Peierls,DJT},
which has become a standard tool in the theory of the QHE \cite{QHE}.

The reduced angular momentum
in the critical case is that of the $\vX_+$-part alone,
\beq
\cJ^c=\cJ_+=\frac{eB_c}{2}\,\widehat{\vX^2_+}=
\frac{\theta}{2}
\,\hat{b}^\dagger\hat{b},
\label{redangmom}
\eeq
consistently with turning off the second term in
(\ref{pmangmom}); it is clearly bounded from one side.

At last, the reduced Hamiltonian is $\hat{V}$ written as
a function of the non-commuting $\widehat{\vX}_+$ alone. In the purely magnetic case, it is thus  \emph{zero}: no motion
is left over.

For a constant electric field, one should
consider plane waves and a continuous spectrum.

For the harmonic trap, it is that of
the $\vX_+$ oscillator alone,
\beq
H^{c}=H_+^{c}=(1+mk\theta^2)\,\frac{k}{2}\,\widehat{\vX_+^2}
=
\theta^2(1+mk\theta^2)\,\frac{k}{2}\,\hat{b}^\dagger\hat{b}\,,
\eeq
which should be compared with the naive expression
$(k/2)\vx^2$.
The reduced energy is hence proportional to the
angular momentum,
\beq
H^{c}=
k\,\theta(1+mk\theta^2)\,\cJ^c.
\eeq

It is worth to mention that for a general (non linear or quadratic) potential $V$ the very quantization procedure is ambiguous \cite{NCLandau}.

\section{Conclusion}

Proceeding backwards, we generalized to non-vanishing
electric fields the remarkable correspondence between the NCLP and
chiral oscillators, advocated before
by Alvarez et al. \cite{AGKP}. Our clue has been replacing the
``natural'' velocity-momentum relation  posited in \cite{AGKP},
\beq
\dot{\vx}=\vp/m,
\eeq
by including the anomalous velocity term as it is usual in the non-commutative context [see
 the first condition in (\ref{exoeqmot})] and,  more importantly,
it is required by the physical applications.
It is the removal of this usual but unjustified
 rule which resolves the problem
 --- just like for anyons \cite{anomalous}.

Details have been worked out for both a constant  electric field and and a harmonic map, providing us with
a quick and elegant derivation of the principal properties of the NCL problem,
cf. \cite{DHexo,NCLandau}.

Particular interest has been devoted to the phase transition which takes place when
the magnetic field crosses the critical value $B_c=(e\theta)^{-1}$, when the effective mass, $m^*$, changes sign. It is  intriguing to investigate
what happens in the ``Hall regime'' $B=B_c$ i.e. $m^*=0$, when only those
motions are consistent which satisfy the Hall law. This is understood by realizing that, when $m^*\to0$, the rotation around the guiding center becomes
more and more rapid --- with the exception, precisely, of those motions which
materialize that of the guiding center. The initial conditions fall hence
into two categories: those which correspond to the Hall law survive
unscathed, whereas the others speed up more and more. Such trajectories become hence more and more dense, becoming ``instantaneous'' as $m^*\to0$.

Do these ``instantaneous motions'' have any physical sense~?
It is hard to say. A footnote in
Ref. \cite{GJ} hints, however, at that they \emph{may}
contribute to the path integral~: according to an unpublished statement of Klauder,  there is finite propagation in infinitesimally short time.

The quantum mechanical description can, just like in classical mechanics \cite{ZH-II}, be derived from the symmetry alone. This is analogous
to the similar derivation of the H-atom spectrum form the
group theory, using the ${\rm O}(4)$ dynamical symmetry.

The transition in the critical case is analogous to the one described in ``Chern-Simons mechanics'' \cite{DJT},
obtained by turning off the kinetic term in the ordinary
Landau problem.

Let us mention, in conclusion, that the chiral
decomposition has recently been extended to the Hill equations of celestial mechanics
 \cite{ZGH-Hill}.


\begin{acknowledgments}
 P.-M. Z. and P.A.H  acknowledge  hospitality at the
 \textit{Laboratoire de Math\'ematiques et de Physique Th\'eorique of Tours University} and at
the \textit{
Institute of Modern Physics in Lanzhou of
the Chinese Academy of Sciences}, respectively. The authors are grateful to C. Duval and
M. Plyushchay  for correspondence.
 This work has been partially supported by the National Natural Science Foundation of
China (Grants No. and 11175215) and by the Chinese Academy of Sciences visiting
professorship for senior international scientists (Grant No. 2010TIJ06).
\end{acknowledgments}
\goodbreak


\end{document}